\documentclass[usenatbib]{mn2e}
\usepackage{graphicx}
\usepackage{graphics}
\usepackage{rotating}
\usepackage{amsmath}
\title[SUMSS II. The
  Source Catalogue]
  {SUMSS: A Wide-Field Radio Imaging Survey of the Southern Sky. II. The
  Source Catalogue}
\author[T. Mauch et al.]
  {T.~Mauch,$^1$\thanks{E-mail: tmauch@physics.usyd.edu.au}
  T.~Murphy,$^1$\thanks{Present address: Institute of Astronomy, University of Edinburgh, Royal Observatory,
  Blackford Hill, Edinburgh EH9 3HJ, UK.}
   H.J.~Buttery,$^2$ 
   J. Curran,$^3$\thanks{Present address: Institute for Communicating
   and Collaborative Systems, School of Informatics, University of
   Edinburgh, 2 Buccleuch Place, Edinburgh, EH8 9LW, UK.}
   R.W.~Hunstead,$^1$
   \newauthor
   B.~Piestrzynski,$^1$ J.G.~Robertson,$^1$ E.M.~Sadler$^1$\\
  $^1$School of Physics,
  University of Sydney, 
  NSW, 2006, Australia.\\
  $^2$Cavendish Laboratory, University of Cambridge, Cambridge CB3 OHE, UK.\\
  $^3$School of Information Technologies, University of Sydney, NSW,
  2006, Australia.\\
  }

\date{Accepted 2003 March 7.}

\pagerange{\pageref{firstpage}--\pageref{lastpage}}

\pubyear{2003}

\begin{document}

\maketitle

\label{firstpage}

\begin{abstract}
This paper is the second in a series describing the Sydney University Molonglo
Sky Survey (SUMSS) being carried out at 843\,MHz 
with the Molonglo Observatory
Synthesis Telescope (MOST). The survey will consist of $\sim590$
\(4.3^\circ\times4.3^\circ\) mosaic images with
\(45''\times45''\rm{cosec}|\delta|\) resolution,
and a source catalogue.
In this paper we describe the initial release (version 1.0) 
of the source catalogue
consisting 
of 107,765 radio sources made by fitting elliptical gaussians in
271 SUMSS
\(4.3^\circ\times4.3^\circ\) mosaics to a limiting
peak brightness of 6\,mJy\,beam\(^{-1}\) at
\(\delta\leq-50^\circ\) and 10~mJy~beam\(^{-1}\) at \(\delta>-50^\circ\).
The catalogue covers
approximately 3500~\(\rm{deg}^2\) of the southern sky with \(\delta\leq-30^\circ\), about 
43 per cent of the total survey area. 
Positions in the
catalogue are accurate to within \(1''-2''\) for sources with peak
brightness
\(A_{843}\geq20\)\,mJy\,beam\(^{-1}\) and are always better than
\(10''\). The internal
flux density scale is accurate to within 3 per cent. Image artefacts
have been classified using a decision tree, which correctly identifies and rejects
spurious sources in over 96 per cent of cases. Analysis of the catalogue shows
that it is highly uniform and is complete to 8\,mJy at
\(\delta\leq-50^\circ\) and 18\,mJy at \(\delta>-50^\circ\). In this release
of the catalogue about 7000 sources are found in the overlap region
with the NRAO VLA Sky Survey (NVSS) at 1.4\,GHz. 
We calculate a median spectral index of \(\alpha=-0.83\) between
1.4\,GHz and 843\,MHz. This version of the 
 catalogue will be 
released via the World Wide Web with future updates
as new mosaics are released.

\end{abstract}


\section{Introduction}

Paper I of this series \citep{bock99} described the 
survey design and science goals of the Sydney University Molonglo Sky 
Survey (SUMSS). SUMSS is imaging the southern ($\delta<-30^\circ$) 
radio sky at 843\,MHz with similar sensitivity and resolution to the 
northern NRAO VLA Sky Survey \citep[NVSS;][]{condon98} at 1.4\,GHz.  

SUMSS uses the Molonglo Observatory Synthesis Telescope 
\citep[MOST;][]{mills81,robertson91}, a 1.6\,km-long cylindrical paraboloid reflector which 
has the largest collecting area of any telescope in the southern hemisphere. 
The MOST was upgraded in 1996--97 to give it a 2.7$^\circ$ diameter field of 
view \citep{large94,bock99}, and since mid-1997 over 90 per cent of 
MOST observing time has been devoted to SUMSS.  The survey will be 
completed by the end of 2003. 

In this paper, we present the first part of the SUMSS source catalogue, 
covering 3500 deg$^2$ of the southern sky.  The catalogue will be updated 
regularly as the survey progresses, and a version is available online 
at www.astrop.physics.usyd.edu.au/sumsscat/.

The structure of the paper is as follows. In Section 2 we describe the
software used to construct the catalogue and the procedures to construct
the source list. In Section 3 we describe our technique for removing
spurious responses from the catalogue. In Section 4 we describe the
uncertainties in the catalogue. Finally, Section 5 contains our analysis  
of the catalogue.

\section{Catalogue Construction}

\begin{figure*}
\includegraphics[width=14cm,angle=180]{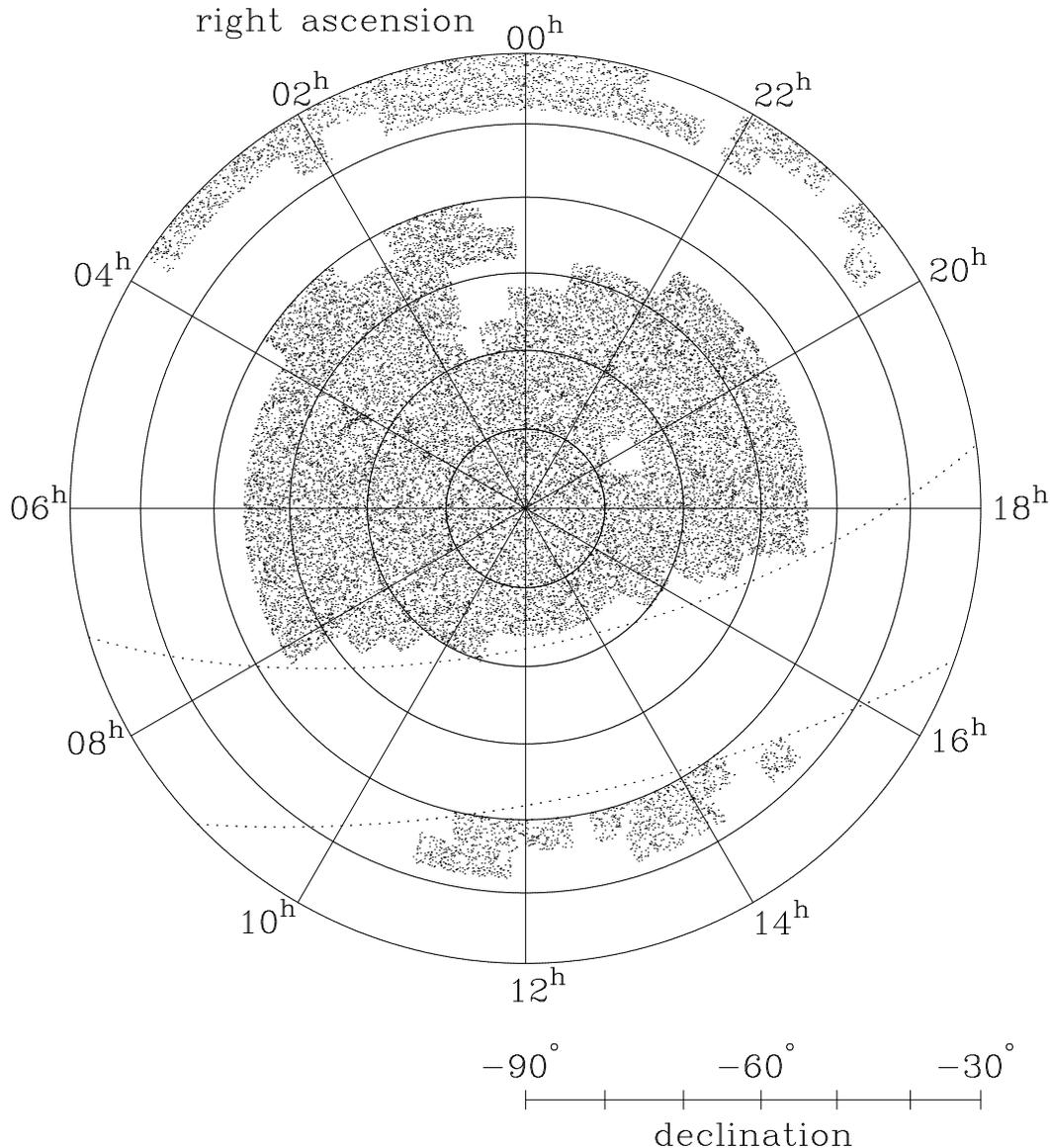}
\medskip
\caption{A plot in the equal-area Lambert projection
  of all 107,765 sources in the 271 mosaics in 
  version 1.0 of the SUMSS catalogue. The total sky coverage is
 3500~deg$^2$.
 Dotted lines are drawn at \(b=\pm10^\circ\) to indicate the
  location of the Galactic plane. 
  The source density is lower at declinations north of
  $-50^{\circ}$ due to the higher flux density limit
  ($S_{843\rm{MHz}}\geq10\,\rm{mJy}$). There is a significant underdensity of
  sources around $\delta=-45^\circ, \alpha=13h$; this is because of the
  large number of artefacts in the vicinity of
  Centaurus A, which obscure weak sources.}
\label{coverage}
\end{figure*}

The individual 2.7$^\circ$ diameter fields of the survey were combined
to form $4.3^{\circ} \times 4.3^{\circ}$ mosaics. The field
centres are located on a grid of overlapping pointing centres such that, 
when combined, sensitivity is recovered in the overlap regions making the
noise in the resulting mosaics almost uniform \citep{bock99}. The mosaic
centres are located on a grid which matches that for the NVSS mosaics, but
is extended to the south celestial pole.
This initial release of the SUMSS catalogue was built on 2002 February 25 
using 271 of the
$\sim590$ mosaics in the complete survey. Figure~\ref{coverage} shows the
positions of all 107,765 sources currently in the catalogue and gives a
representation of the sky coverage of this release. 

Extracting sources from astronomical images is a well documented problem
and there are currently
many computer programs which will find and
characterise sources in images such as those in SUMSS. We decided to use
the {\sc aips} task {\sc vsad}, written for the NVSS survey~\citep{condon98},
which locates
sources in an image and fits elliptical gaussians to them. This was to ensure
 uniformity between SUMSS and NVSS, and also because our tests showed that 
{\sc vsad} fitted sources more reliably than other programs such as {\sc
  imsad} in the {\sc miriad} package \citep{sault95}.

Most of the sources in the SUMSS survey are well fitted by an elliptical
gaussian model because the majority of extragalactic radio sources are smaller
than the MOST restoring beam of $45''\times45''\rm{cosec}|\delta|
$ \citep{windhorst1990}.
The current release of the SUMSS catalogue does not cover the Molonglo Galactic Plane
Survey (MGPS-2; \citealt{green99}) region
($|b|<10^\circ$) because complex
source structures in the Galactic plane make elliptical
gaussian fits unsatisfactory. However, 
there is little
contamination by complex Galactic sources in MGPS-2 mosaics as close to the
Galactic plane as $|b|=2^\circ$, so in the future it will be possible to
visually inspect
regions closer to the Galactic plane to decide those which can be included 
in the catalogue using current methods.
In extremely complex regions we intend to 
crossmatch the MGPS-2 mosaics with source catalogues
at other frequencies (eg. IRAS\,PSC; \citealt{iraspsc} \& RASS;
  \citealt{RASS}).

\begin{figure*}
\includegraphics[width=9cm]{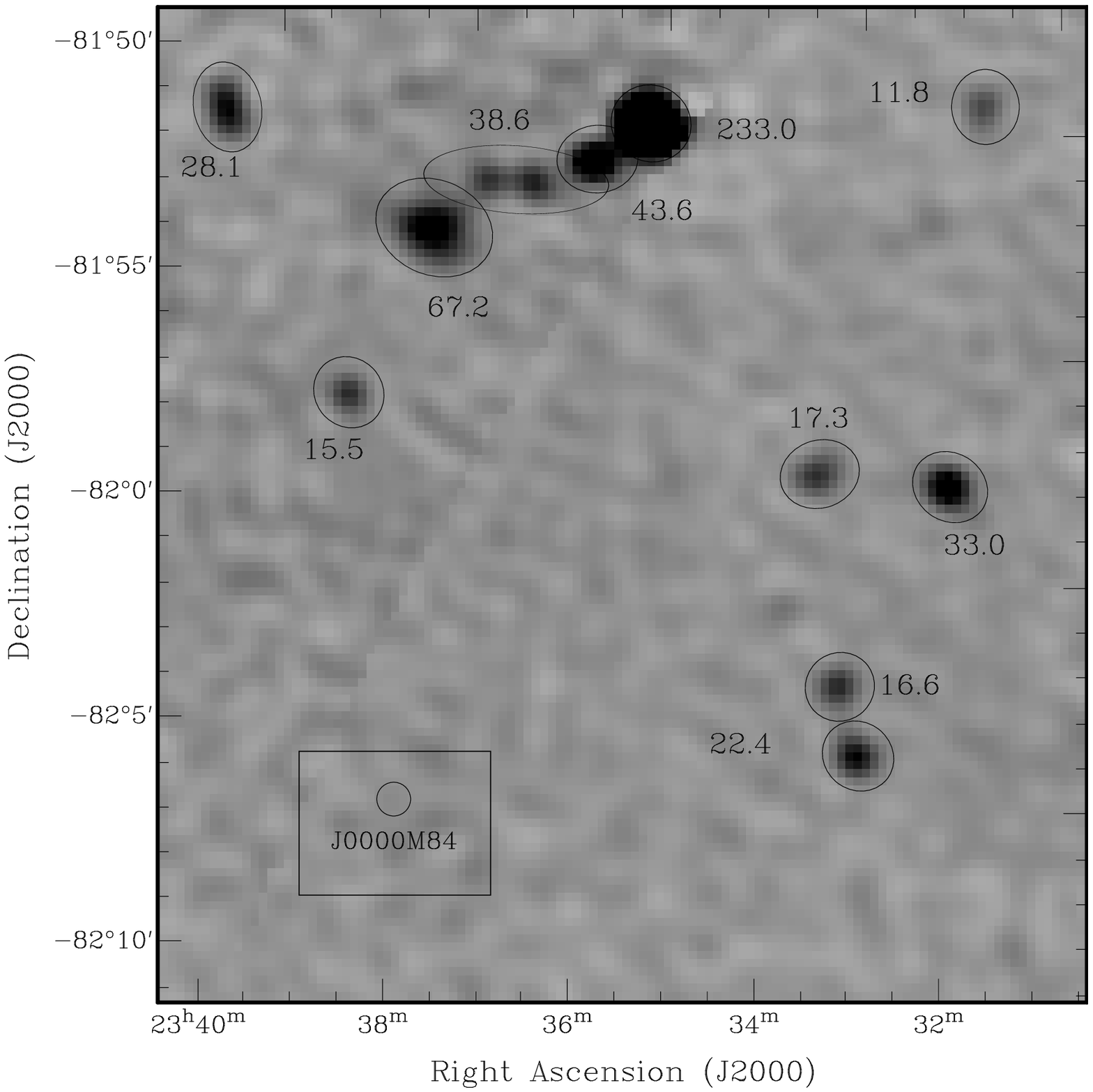}%
\includegraphics[width=9cm]{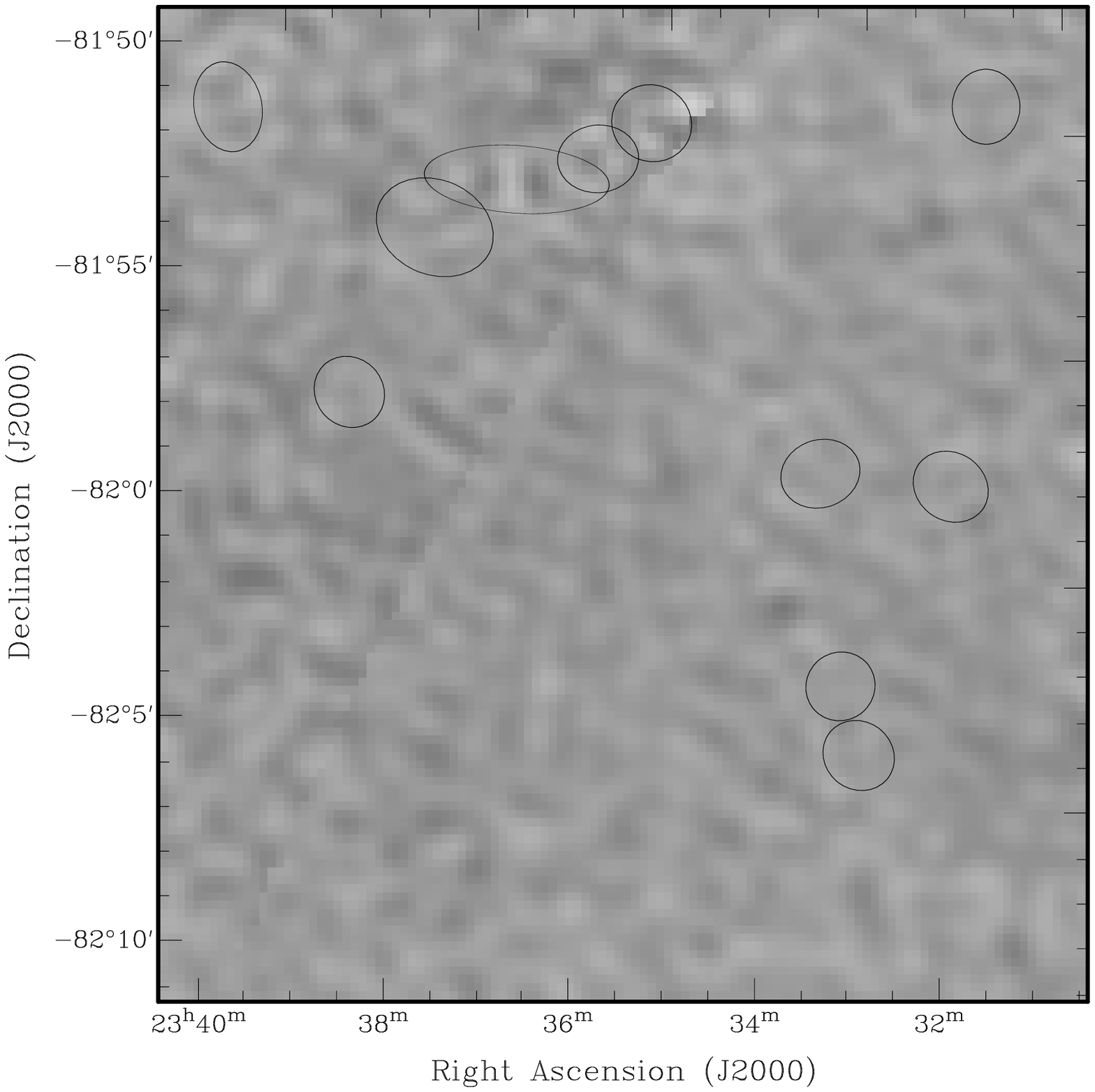}
\caption{Left: A small section of the mosaic J0000M84 with ellipses fitted 
  by {\sc vsad}. The total flux density 
  (in mJy) of each source is printed beside
  it. 
  The beam is shown as a small circle on the bottom left of
  the image. One close double is fitted as a single
  gaussian, while another is fitted as two. Right: The residual image of
  the same region after subtraction of the fitted
  gaussians. The improperly fitted
  source in the north has resulted in residual flux in this image.}
\label{fitting}
\end{figure*}

\subsection{Source Fitting}
\label{sourcefitting}

{\sc vsad} was used to find radio sources
in each of the $4.3^{\circ} \times 4.3^{\circ}$ mosaics in
the SUMSS survey and fit an elliptical gaussian to them. 
The parameters of each gaussian returned by {\sc vsad} are the
J2000 right ascension $\alpha$ and declination 
$\delta$ (both in degrees), peak brightness 
$A_{843}$ (mJy beam$^{-1}$), total flux density
$S_{843}$ (mJy), FWHM fitted source major and minor axes
$\theta_{\rm{M}},\theta_{\rm{m}}$  
(arcseconds)
and the fitted position angle of the major axis (degrees east from north). 
{\sc vsad} also
creates a residual image by subtracting each fitted gaussian from the input
image. 

Figure~\ref{fitting} shows a small region of an illustrative SUMSS mosaic
(J0000M84\footnote{The naming scheme for SUMSS mosaics is J$hhmm$M$dd$ where J
  signifies J2000 coordinates, $hhmm$ is the RA in hours and minutes of the
  mosaic centre, M signifies southern declination and $dd$ is the declination
  of the mosaic centre in degrees.}) 
overlaid with the gaussians fitted by {\sc vsad}.
Most sources on the original mosaic are well fitted by gaussians, though
some artefacts close to stronger sources are also fitted. 
It can be seen from the complex structure in
Figure~\ref{fitting} that {\sc vsad} can be
unreliable for close pairs of 
sources in extreme cases. 
Occasionally two distinct sources
are fitted incorrectly as a single
gaussian with major axis greater than the true separation of the
sources (eg. the 38.6\,mJy extended source in
Figure~\ref{fitting}.). Almost all
close doubles remain in the final version of the catalogue as separate
sources.

\begin{figure}
\includegraphics[width=\linewidth]{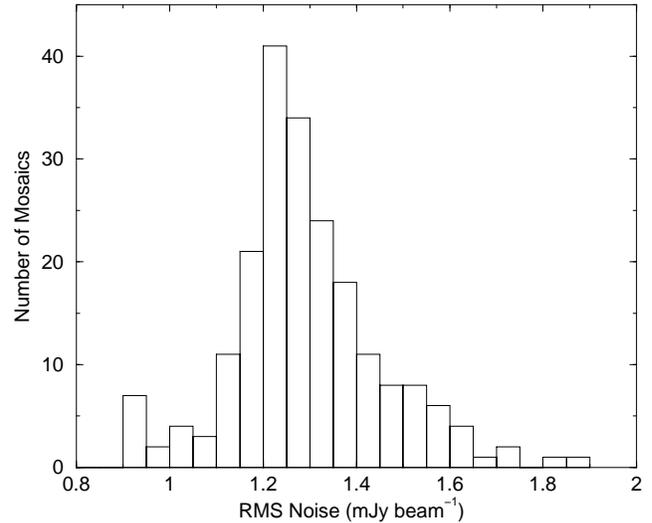}
\caption{A histogram of the rms noise measured in the mosaics at
  $\delta\leq-50^{\circ}$. Most of the mosaics have rms values 
  $\sim1.2-1.3$ mJy beam$^{-1}$; the mode is 1.25 mJy beam$^{-1}$. 
  The tail at higher rms values is due to
  mosaics containing bright sources which tend to increase the
  local rms noise.} 
\label{noisehisto}
\end{figure}

\subsection[]{Noise}

\begin{figure}
\includegraphics[width=\linewidth]{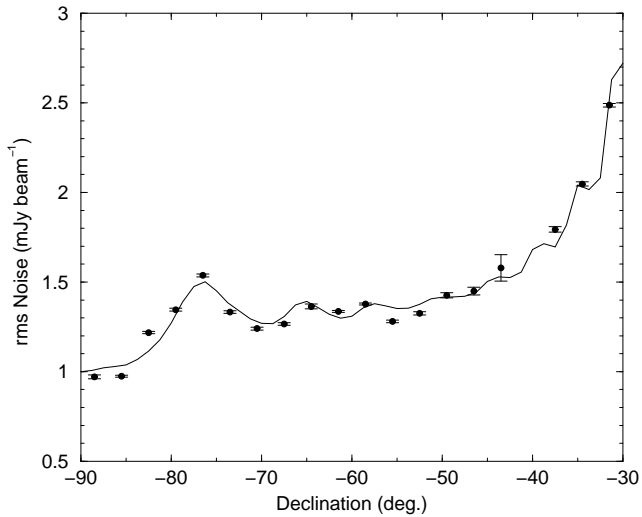}
\caption{A plot of median rms noise computed around brighter sources
  $(S_{843}>50\,\rm{mJy})$ vs. declination.  
  The line drawn is a model based on the MOST MD gain curve. 
  It is shown to indicate
  the effect of the variation of the MOST gain with declination. The noise
  peak at $\delta=-76^\circ$ is explained by this curve. The
  rms noise increases sharply north of
  $\delta\sim-50^{\circ}$, which is the declination above which we increase
  the brightness
  limit of the catalogue to $10\,\rm{mJy}\,\rm{beam^{-1}}$ .}
\label{rmsdec}
\end{figure}

Even though the field tiling patterns were designed to make the resulting noise in
the mosaics uniform, background noise is higher
in localised regions close to strong sources.
We determined the rms noise in the SUMSS survey to 
establish a
threshold below which sources are discarded from the catalogue.
The residual images created by {\sc vsad} were kept and used to estimate the local 
rms noise for each source. Two estimates of the rms noise were obtained: 

\begin{enumerate}
\item First
the noise was estimated over each residual image by
fitting a normal curve to the pixel distribution giving
an estimate of the average noise over the area of each entire mosaic. 
Figure~\ref{noisehisto} shows the distribution of rms values
obtained in this way. The median rms noise 
of the mosaics at $\delta\leq-50^{\circ}$ is 1.27\,mJy\,beam$^{-1}$. For $\delta>-50^{\circ}$
  the scatter is much greater as the rms 
  noise increases strongly with declination north of $-50^\circ$ (see
  Figure~\ref{rmsdec}). The median value of rms noise at
  $\delta>-50^{\circ}$ is 1.9\,mJy\,beam$^{-1}$.  Hereafter we refer to regions at
$\delta\leq-50^{\circ}$ as southern and those $\delta>-50^{\circ}$ as
northern. 

\item Secondly, a local rms for each source was determined by computing a 
pixel histogram in a box of $100\times100$ pixels ($\sim$600 beams) 
in the residual image and
fitting a normal curve to this distribution. The 
distribution was clipped so as to
only include pixels within $\pm5$ times the rms noise measured
in the mosaics. Residual
artefacts from strong sources
tend to increase the estimate of local rms
noise. By this method we have an
estimate of the local rms noise for each source which takes into
account increases in the noise level close to bright sources.
Values of local rms noise close to bright sources are
3--4 times greater than those in other regions, because of
the limited dynamic range $(\sim100:1)$ of the MOST \citep{bock99}.  
\end{enumerate}

Figure~\ref{rmsdec} shows the variation in local rms noise with
declination for stronger sources $(S_{843}>50\,\rm{mJy})$ in the present survey
release. The gain of the MOST varies with Meridian Distance\footnote{The Meridian
  Distance (MD) at declination \(\delta\) and hour angle \(H\) is given 
  by \(\sin({\rm MD}) = \cos \delta \sin H\). This is explained in more
  detail in Paper I \citep{bock99}.} (MD) due to a
number of factors arising from the structure of the telescope. For example
the $\rm{cosec}|\delta|$-shaped rise towards 
northern declinations is due to the foreshortening of the
MOST at large $|\rm{MD}|$.
A model of the rms noise variation with declination 
is plotted. This was determined
by summing the noise variance as a function of hour angle, taking into account
the variation of noise due to the MOST MD gain curve.
This curve matches the observed rms noise variation quite well.  

{\sc vsad} was used to fit all peaks brighter than 5 mJy beam$^{-1}$ at
$\delta\leq-50^{\circ}$ and
10 mJy beam$^{-1}$ at $\delta>-50^{\circ}$. Typically
about 400 gaussians were fitted in the northern mosaics and about 700
in the southern ones. {\sc vsad} fitted many noise peaks and 
artefacts in the southern mosaics 
between 5 and 6\,mJy\,beam$^{-1}$. 
We decided to set the catalogue limit at 
6\,mJy\,beam$^{-1}$ at
$\delta\leq-50^{\circ}$ and 10\,mJy\,beam$^{-1}$ at
$\delta>-50^{\circ}$.

\subsection{Duplicate Sources}
\label{duplicates}

The {\sc vsad} routine was run separately on each mosaic
resulting in a total list of 171,846 responses over the
3500 deg$^2$ of survey area currently complete. This list was
then pruned to remove multiple entries and spurious sources arising from
image artefacts.
The SUMSS survey was designed such that the mosaics created from the
individual observations overlapped slightly \citep{bock99}. 
This overlap changes with
declination from about 50 per cent at $\delta=-88^{\circ}$ to 1--2 per cent at
$\delta=-32^{\circ}$.
Therefore some entries in the total list are 
multiple occurrences of sources which have 
appeared in overlapping mosaics.

\begin{figure}
\includegraphics[height=\linewidth,angle=-90]{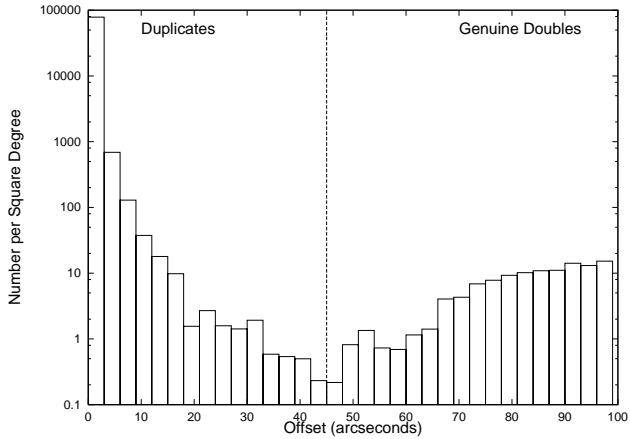}
\caption{The distribution of source separations   
  for the raw version of the SUMSS catalogue. 
  The number of sources found at different radii has been divided by the
  sky area to obtain the source density versus separation.
  The minimum at
  about 45$''$ is taken as the separation 
  beyond which the number of close doubles
  dominates over the number of duplicates. There is a small number of
  sources with separations between $45''$ and $75''$; these are the result
  of overlapping gaussians fitted by {\sc vsad}. Above $80''$ the
  distribution flattens as expected.}
\label{dups}
\end{figure}

Figure~\ref{dups} is a plot of the distribution of source separations 
for a subset of about 40,000 sources in the raw
catalogue. The distribution has a minimum at 45$''$ after which the
contribution of genuine close doubles causes it to rise
again. This minimum is not surprising given that the beamwidth of the survey
is $\sim45''$. Sources appearing in different mosaics with 
position differences
 less than
45$''$ were flagged as possible duplicates.

Once a group of duplicate sources has been identified the following
criteria are used to select which source to retain from that group:

\begin{enumerate}

\item If there are more than two sources then the peak amplitudes of all
  the sources are compared and those
  with peak amplitude greater than twice or less than half the average are
  ignored. 
  This can occur
  when some artefacts are fitted at the position of another source.

\item Sources closer than 10 pixels from the edge of a mosaic are ignored to
  ensure that extended sources are not fitted over the edge of the
  image. If all sources are further than 10
  pixels from the edge then option (iii) is used.

\item If there is still more than one source to select, the source with the
  lowest local rms noise has its fitted parameters recorded in the catalogue.

\end{enumerate}

These criteria ensure that one source is selected out of a set of multiple
detections. 
Generally only two sources need to be compared and the one with
the lower local rms noise is placed in the catalogue. A flag telling how
many other mosaics the selected source appears in is included in the final
catalogue. The source which is listed in the catalogue can be considered the
best-fitting source from {\sc vsad}. In compiling the current version of the catalogue, about 45,000 
sources were removed because 
they appear more than once in the raw catalogue.

\section{Image Artefacts}

\begin{figure*}
\includegraphics[width=5.8cm]{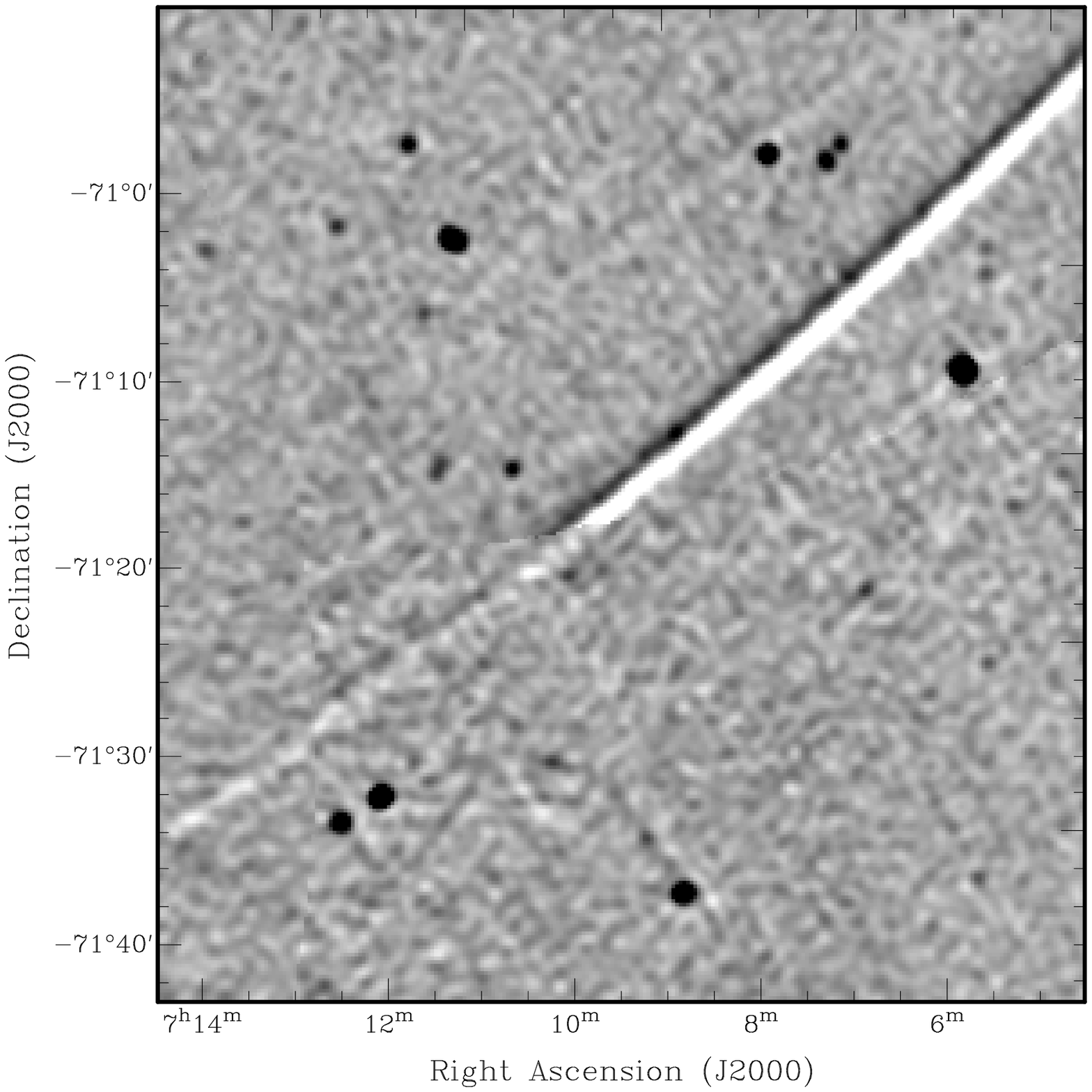}%
\includegraphics[width=5.8cm]{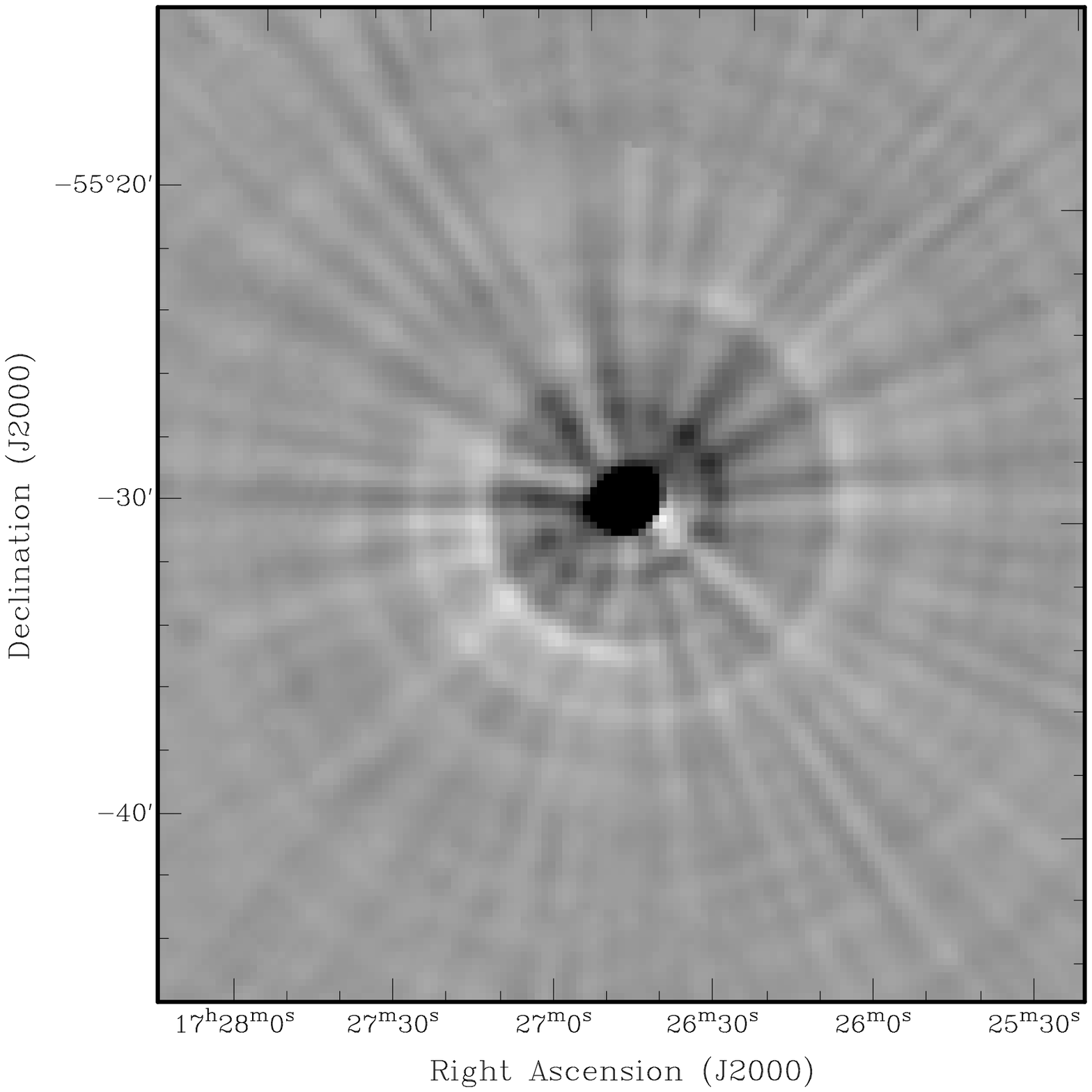}%
\includegraphics[width=5.8cm]{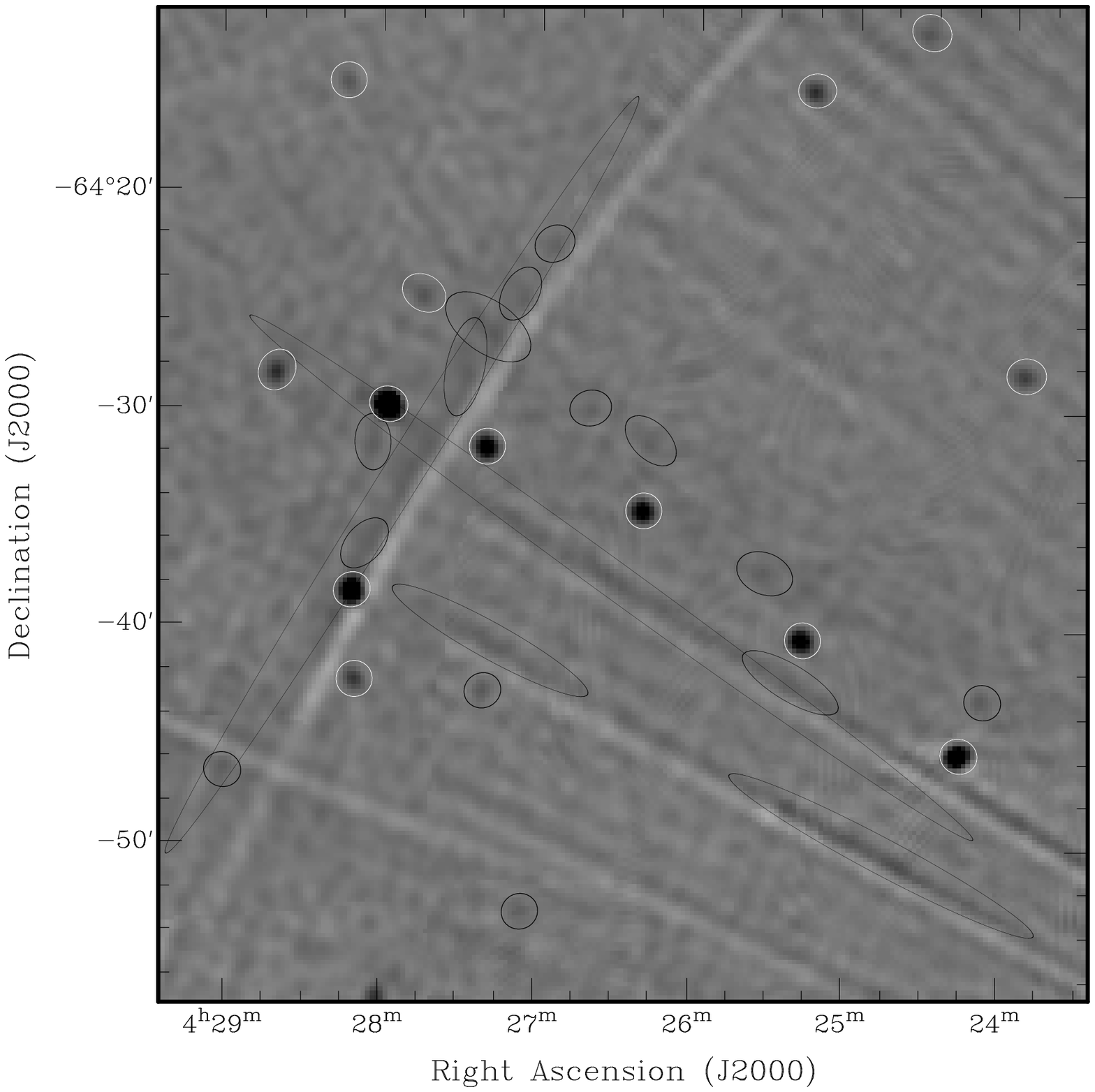}
\caption{Three images showing the variety of artefacts present in SUMSS
  mosaics which are fitted as spurious sources by {\sc vsad}. The left image
  shows part of a grating ring associated with a 5.46 Jy source
  (PKS\,B0743--673) at
  $\rm{RA}=07^{\rm{h}}\,43^{\rm{m}}\,32\fs67$,
  $\rm{Dec.}=-67^\circ\,26'\,28\farcs4$ (J2000). Its
  sudden weakening occurs because the ring appears in different parts of
  two 2.7$^\circ$ fields which make up the mosaic. The middle
  image shows a 1.8 Jy source and the radial spikes associated with
  it. The spikes here have peak amplitude of 10 mJy beam\(^{-1}\)
  and extend about
  30$'$ away from the source. The right image shows 
  a small area of the SUMSS mosaic J2100M72 with many artefacts. 
  Radial spokes of average peak amplitude $\sim6$
  mJy beam$^{-1}$ are visible, as is a weak grating ring. 
  Ellipses fitted by {\sc vsad} are overlaid on the
  image. Sources classified by the decision tree as artefacts are shown as
  black ellipses and sources classified as genuine are shown as white ellipses. }
\label{artifacts}
\end{figure*}

The images in SUMSS are affected by a number of artefacts, many of which
are fitted by the {\sc vsad} routine as sources.
Classification of artefacts is a difficult problem, as they vary
enough in shape and strength that no simple method can be used to separate them from
real sources. The images in 
Figure~\ref{artifacts} show the variety of artefacts in SUMSS
images and some examples of elliptical gaussians which 
{\sc vsad} has fitted to them. About 10 per cent of sources in the raw
SUMSS catalogue are fits to artefacts.

\subsection{Types of Artefacts}

There are two common artefacts which are fitted as genuine sources by
{\sc vsad}, \textit{grating rings} and \textit{radial spokes}.

\textit{Grating rings} arise from the periodic structure of the MOST
array. They appear as ellipses of semi-diameter $n(1.15^\circ
\times 1.15^{\circ}\rm{cosec}|\delta|)$ where $n$ is an integer denoting the
order of the grating ring. Grating rings increase in strength up to the
fourth order ring. While grating
rings are not uniquely a problem of the MOST, it is not possible to remove
them using the standard CLEAN routine. This is because the effective
primary beam of the MOST varies with time and is not azimuthally symmetric.
Also individual baseline visibilities are not recorded during MOST observations
\citep{mills81,robertson91}. 
It is possible that in future an improved CLEAN routine might be written to
remove grating rings by accurately modelling the MOST primary
beam. The future upgrade of the telescope as an SKA demonstrator
\citep[SKAMP;][]{ska}
will eliminate this
problem, as all the baseline visibilities will be retained with each
observation and application of
self calibration algorithms will be possible.
The amplitude of
grating rings is also dependent on the position of the source
in the original SUMSS field so the
mosaicing process can cause abrupt steps in their strength. This effect
can be seen in the leftmost image of
Figure~\ref{artifacts}, where a strong fourth-order 
grating ring associated with a 5.4 Jy source abruptly
becomes weaker. 

\textit{Radial spokes} appear as long thin bands stretching away radially
from a source. They are believed to be caused by random shifts of order
1$''$ in the position of the comb of MOST fan beams, on
a time-scale of minutes \citep{bock99}. The shifts are believed to
occur due to weather related
anomalies in the local oscillator phase or irregularities in ionospheric
or tropospheric refraction. Some radial spokes are shown in the middle
image of Figure~\ref{artifacts} and almost always appear as highly
elongated
 sources in the
output of {\sc vsad}. Generally such artefacts are only fitted close to
sources with peak brightness greater than 500\,mJy\,beam$^{-1}$, because 
the peak amplitude of radial
spokes then becomes stronger than 6\,mJy\,beam$^{-1}$. 
The source density close
to extremely bright sources in SUMSS can decrease substantially as radial
spokes can raise the local rms noise level by up to
3-4 times.

\subsection{Response Classification with a Decision Tree}

\begin{table}

\caption{Confusion matrices comparing decision tree classification with hand
  classification. Left: Results of southern decision
  tree on test data. 
  Right: Results of northern decision tree on test data.}
\label{classresults}
\begin{center}
\begin{tabular}{@{}rccc}
\multicolumn{4}{c}{Southern}\\
\hline
Class & \multicolumn{3}{c}{(decision tree)}\\
(hand)& (1) & (2) & (3) \\
\hline
(1) & 12 &   & 5 \\
(2) & 1 & 2 &  \\
(3) & 1 & 3 & 618 \\
\end{tabular}
\begin{tabular}{@{}rccc}
\multicolumn{4}{c}{Northern}\\
\hline
Class & \multicolumn{3}{c}{(decision tree)}\\
(hand)& (1) & (2) & (3) \\
\hline
(1) & 3 & 1 & 1 \\
(2) & 2 & 1 &  \\
(3) & 3 &  & 415 \\
\end{tabular}
\end{center}
\end{table}

The variety and complexity of artefacts fitted as sources make it difficult
to classify sources as genuine or spurious in a simple way.
We have employed the \textit{decision tree} program
{\sc c4.5}~\citep{Quinlan93} to aid in the classification of sources in the mosaics.
A decision tree encodes a classification function as a
hierarchy of tests which classify examples into different classes.
Each example consists of a set of attributes, each with an associated value.

The tree represents the classification function as follows:

\begin{enumerate}
\item Every internal node corresponds to a test examining one or more
attributes of the example to be classified.
\item Each branch descending from the node corresponds to a particular outcome
of the test.
\item Finally, the leaf nodes in the tree are labelled with the class to assign. 
\end{enumerate}

{\sc c4.5} is a popular, freely available decision tree learner. Another application
of a decision tree to radio source catalogues can be found in the FIRST
survey \citep{white97}.

%Each new example is classified by recursively selecting a branch
%to follow, based on the outcome of the test at the current node,
%starting from the root node.
%When a leaf node is reached, the label of the leaf node is assigned

%as the class of the example.

%

%A decision tree learner induces the tree structure from a hand classified training set.

%C4.5 is a popular, freely available decision tree learner.

%Each internal node in a decision tree partitions the training set into subsets

%based on the outcome of the test.

%C4.5 uses the \textit{GainRatio} measure, based on the entropy of these subsets,

%to select the most informative test at each node in the tree.

%After the tree is constructed, nodes supported by little evidence

%from the training set are pruned to reduce overfitting. Another application

%of a decision tree to radio source catalogues can be found in the FIRST

%survey \citep{white97}.

The raw output from {\sc vsad} includes about 39 source characteristics which can be
used as attributes for the decision tree; these include the fitted
and deconvolved
source sizes and the raw fitting uncertainties 
quoted by {\sc vsad}. We have also included some extra
parameters mainly relating each fitted source to the nearest strong source, including 
separation and relative position angle. 
Because of the different properties of the MOST beam
in the two regions, we made two separate
decision trees, one for the southern 
catalogue ${\delta}\leq-50^{\circ}$ and one for
the northern catalogue
${\delta}>-50^{\circ}$.

The decision tree program was trained by hand on a subset of mosaics with
many artefacts. About 3000 sources in the south and about 1500 sources in
the north were hand classified. To ensure that the training was reliable,
each source was examined separately
by at least two of us, and the final decision on each source was made by one
of us (Mauch) comparing the two human classifications, thereby minimising 
the subjective judgement of each human classifier. About 10 per cent of
human classifications were changed during comparison. 
The objects contained in the raw output of {\sc vsad}
were classified using a numbering scheme from 1-3 with the following definitions. 

\begin{enumerate}
 \renewcommand{\theenumi}{(\arabic{enumi})}
\item The source is an artefact. 
\item The source is in a region of low signal-to-noise.
\item The source is real.
\end{enumerate}

The decision tree was then run and the result for both the northern
and southern decision trees was tested on a small independent 
sample of hand classified
sources from one mosaic. Table~\ref{classresults} shows the results from
this test. It can be seen from 
analysis of the training data that artefacts tend to be correctly
identified.
In some
cases a hand classified type 1 source is machine classified as type 2 and vice versa but it is
rare for sources hand classified as 1 or 2 to be machine classified as
genuine. 
Conversely
only a small number of type 3 sources have been classified as artefacts. This does have an effect
on the completeness of the survey at flux densities of 6--10\,mJy (Figures~\ref{density}~\&~\ref{lognlogs} in
Section~\ref{analysis} show this more clearly). 
Overall the accuracy of the decision tree on 
the testing data is conservatively estimated to be 96 per cent. 
The accuracy for the entire catalogue is
probably a little lower, but most of the misclassified sources are real
sources which were classified as artefacts.
This implies that the final version of the catalogue 
should be very reliable. 

The decision tree was found to be least reliable for extended sources
with major axis length greater than five times their minor axis
length. Every such source classified as genuine by the decision tree was
double checked by hand. About 10 per cent
of these classifications were found to
be incorrect. 

The final released catalogue only
contains sources classified as genuine by the decision tree. The
unmodified
catalogue 
contains 10 per cent extra sources, most of which are artefacts.  This
larger (but less reliable)
catalogue with an extra column containing decision tree classifications, 
will be available on request.

\section{Accuracy}

All of the uncertainties calculated in the catalogue are a combination of both
fitting and calibration uncertainties of the MOST. In general
the calibration uncertainties of the MOST are small and the fitting uncertainties tend to
dominate. Fitting uncertainties for the SUMSS catalogue were determined using equations derived
in~\citet{c97} (herafter C97). The noise in SUMSS mosaics is correlated at the length
scale of the restoring beam of MOST. As this beam is elliptical, the
axis length of the restoring beam was taken as the noise correlation
length in SUMSS
($\theta_{\rm{N}}^2=b_{\rm{M}}b_{\rm{m}}=45\arcsec\rm{cosec}|\delta|\times45\arcsec$;
where $b_{\rm{M}},b_{\rm{m}}$ are widths of the MOST beam major and minor 
axes respectively). This implies that the effective
signal-to-noise ratio ($\rho$) is given by:
\begin{equation}
\rho^2 = \frac{\theta_{\rm{M}}\theta_{\rm{m}}}{4\theta_{\rm{N}}^{2}}\left[1 + \left(\frac{\theta_{\rm{N}}}{\theta_{\rm{M}}}\right)^2\right]^{\alpha_{\rm{M}}}\left[1 + \left(\frac{\theta_{\rm{N}}}{\theta_{\rm{m}}}\right)^2\right]^{\alpha_{\rm{m}}}\frac{A_{843}^2}{\sigma^2},
\label{sn}
\end{equation}
where $\theta_{\rm{M}}$ is the fitted major axis size and $\theta_{\rm{m}}$ is the fitted
minor axis size. $A_{843}$ is the peak brightness
of the fitted Gaussian and $\sigma$ is
the rms noise of each mosaic (C97); $\sigma$ is taken to be the local
rms noise as derived in Section~\ref{sourcefitting}. For the uncertainties
in each fitted parameter
we have used the same empirical values for the exponents
$\alpha_{\rm{M}}$ and $\alpha_{\rm{m}}$ taken from C97 
and also used in the NVSS source
catalogue~\citep{condon98}.

\subsection{Position Uncertainties}

The fitting variances in the source positions are given by:
\begin{equation}
\sigma{_\alpha^2}=\sigma{_{\rm{M}}^2}\sin^2\left({\rm
    PA}_{\rm F}\right)+\sigma{_{\rm{m}}^2}\cos^2\left({\rm PA}_{\rm F}\right),
\label{raerr}
\end{equation}
\begin{equation}
\sigma{_\delta^2}=\sigma{_{\rm{M}}^2}\cos^2\left({\rm
    PA}_{\rm F}\right)+\sigma{_{\rm{m}}^2}\sin^2\left({\rm PA}_{\rm F}\right),
\label{decerr}
\end{equation}
where the rms noiselike uncertainties of the fitted major and minor axes
$\sigma{_{\rm{M}}}$ and $\sigma{_{\rm{m}}}$ are derived as in equation~25
of C97. PA$_{\rm F}$ is the fitted position angle of the source in degrees
east of north. The fitting uncertainties become quite large ($\sim5\arcsec$) for
weaker extended sources $(A_{843}<10\,\rm{mJy}\,\rm{beam}^{-1})$. 

\begin{figure}
\includegraphics[width=\linewidth,height=8.4cm]{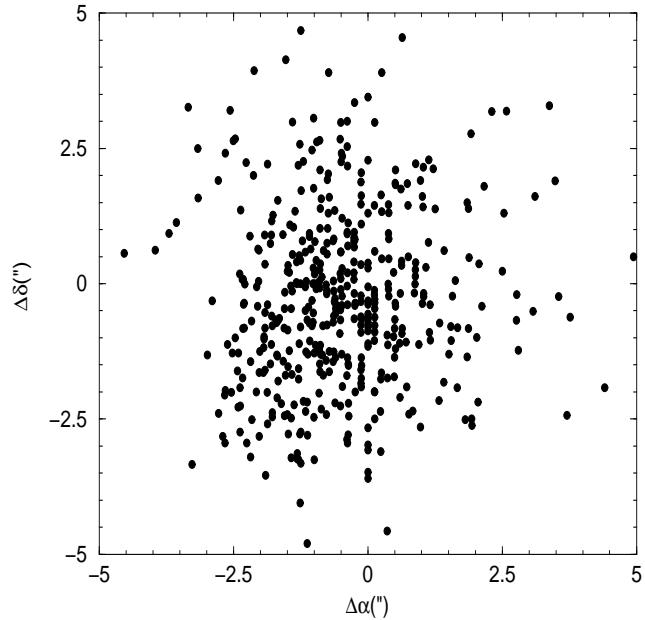}
\caption{A plot of the offset (SUMSS position minus NVSS position)
  in both Right Ascension $(\alpha)$ 
  and Declination $(\delta)$ between 500 bright point sources
  in the overlap region between SUMSS and NVSS.}
\label{brightoffsets}
\end{figure}

For sources with $S_{843}\geq50\,\rm{mJy}$
the calibration 
uncertainty of the MOST is greater than the fitting
uncertainties. We have determined the calibration
uncertainty for stronger sources $\left(S_{843}>200\,\rm{mJy}\right)$
by comparison with positions in the
NVSS catalogue in the overlap region between SUMSS and NVSS
($-40^{\circ}<\delta<-30^{\circ};\,464\,\rm{deg^2}$ in the current
release). 
The positions of strong sources in the
NVSS catalogue are known to be accurate to within
($\epsilon_{\alpha},\epsilon_{\delta}$)=($0\farcs45,0\farcs56$)
\citep{condon98}.
Only point sources in SUMSS were used in this comparison to avoid
the larger position uncertainties associated
with extended
sources. There are about 500 SUMSS sources which meet these criteria in the
overlap region, all of which have a counterpart in the NVSS
catalogue. Figure~\ref{brightoffsets} contains a plot of the offsets
in Right Ascension ($\Delta\alpha$) and
Declination ($\Delta\delta$) between
NVSS and SUMSS. The mean offsets are $\left<\Delta\alpha\right>=-0\farcs59\pm0\farcs07$ and
 $\left<\Delta\delta\right>=-0\farcs30 \pm 0\farcs08$. These offsets are
 caused by a combination of both calibration errors in individual fields
 and fitting errors. No correction has
 been applied to the final version of the catalogue. 
For more information about SUMSS position uncertainties, see Paper I \citep{bock99}.

The rms of the
offsets between the SUMSS and NVSS catalogues have been used to determine the
SUMSS position calibration uncertainties. These are
$\epsilon_{\alpha}=1\farcs5$ for Right Ascension and
$\epsilon_{\delta}=1\farcs7$ for Declination. The elongation of the MOST
beam at $\delta\geq-40^\circ$ means the rms in declination is larger than
in the southern catalogue, implying that these values are a
conservative estimate of the MOST position uncertainty in declination.
The calibration uncertainties quoted here have been added
in quadrature to the fitting uncertainties for each source to obtain
the position uncertainties in the catalogue.

%\begin{figure}
%\includegraphics[width=\linewidth,height=\linewidth]{sumssatca.eps}
%\caption{The offsets between SUMSS and the ATCA for a selection of ??
%  sources observed with both surveys using the 6A configuration of the
%  AT, the offsets are normalised by the SUMSS position errors. 
%  A 3$\sigma$ error circle is also plotted. The errors appear smaller in
%  declination because our calibration error term was estimated where the
%  beam was elongated.}
%\label{faintoffsets}
%\end{figure}

\subsection{Source Sizes}

The uncertainties in the axes of the elliptical gaussians fitted by {\sc vsad}
\(\left(\sigma(\theta_{\rm{M}}),\sigma(\theta_{\rm{m}})\right)\) 
are determined by combining in quadrature
equation 21 of C97 and the
calibration uncertainty in the major and minor axes of the MOST beam
shape. The MOST beam calibration
uncertainty has been determined by examining fits to moderately strong
sources believed to be unresolved. Because fits to the strongest
sources with $S_{843}>500\,\rm{mJy}$ could be contaminated by image
artefacts, only moderately strong
(100\,mJy$<S_{843}<$500\,mJy) sources were chosen. From this
analysis we have conservatively 
estimated the beam calibration uncertainty to be
$\epsilon_\theta=3$ per cent in both axes. The uncertainty
is worst in the northern mosaics in
which the beam is considerably elongated.
The
probability that the fitted size of a point source would be larger
than the beam by more than \(2.33\sigma(\theta_{\rm{M,m}})\) 
is 2 per cent so we compare the beam
plus $2.33\sigma(\theta_{\rm{M,m}})$ with
 the major and minor fitted axis lengths to determine
if a source is resolved along either axis. 

Sources for which either the major axis or both fitted axes are believed to
be resolved are then deconvolved along each resolved axis.
The fitted gaussians in the raw output of {\sc vsad} are the convolution of the
MOST elliptical beam with the true source shape. The deconvolved major and minor axis widths of 
each fitted source \((\phi_{\rm{M}},\phi_{\rm{m}})\) were found using
\begin{equation}
2\phi_{\rm{M}}^2 = (\theta{_{\rm{M}}^2} + \theta{_{\rm{m}}^2}) - (b{_{\rm{M}}^2} + b{_{\rm{m}}^2}) + \beta,
\end{equation}
\begin{equation}
2\phi_{\rm{m}}^2 = (\theta{_{\rm{M}}^2} + \theta{_{\rm{m}}^2}) - (b{_{\rm{M}}^2} + b{_{\rm{m}}^2}) - \beta,
\end{equation}
where \(\theta_{\rm{M}},\theta_{\rm{m}}\) are the fitted major \& minor
axes of the source, 
\(b_{\rm{M}},b_{\rm{m}}\) are the beam major \& minor axes and \(\beta\) 
is given by
{\setlength\arraycolsep{2pt}
\begin{eqnarray}
\beta^2 & = & {({\theta_{\rm{M}}^2}-{\theta_{\rm{m}}^2})^2}+{(b_{\rm{M}}^2 -
  b_{\rm{m}}^2)^2}-{}\nonumber\\ 
& &
  2(\theta_{\rm{M}}^2-\theta_{\rm{m}}^2)(b_{\rm{M}}^2-b_{\rm{m}}^2)\cos2\left({\rm PA}_{\rm{F}} - {\rm PA}_{\rm{B}}\right),
\end{eqnarray}
where \({\rm PA}_{\rm{F}}\) and \({\rm PA}_{\rm{B}}\) are the position angles of the fitted
source and the MOST beam~\citep{wild70}. The MOST beam is oriented north-south
so \({\rm PA}_{\rm{B}}=0\) always.

The ellipticity of the MOST beam causes the fitted position angle to differ
from the true source position angle (PA$_{\rm S}$). We find the deconvolved 
major axis position angle using
\begin{equation}
\tan{(2\rm{PA}_{\rm{S}})}=\left[\frac{(\theta{_{\rm{M}}^2} -
      \theta{_{\rm{m}}^2})\sin{2\rm{PA}_{\rm{F}}}}{(\theta{_{\rm{M}}^2} -
      \theta{_{\rm{m}}^2})\cos{2\rm{PA}_{\rm{F}}}-{(b_{\rm{M}}^2-b_{\rm{m}}^2)}}\right].
\end{equation}

A deconvolved 
source size is quoted for each resolved source in the catalogue. No
source sizes are given for unresolved sources. It should be noted that the
fitted source sizes and deconvolved source sizes in the catalogue are only
intended to be indicative of the true source structure. Sometimes more extended
sources are the result of a poor fit by the {\sc vsad} program (see
Figure~\ref{fitting}) and the original
images are the best guide in determining whether or not a given radio
source is resolved.

\begin{table*}
\begin{minipage}{145mm}
\caption{Measurements of Molonglo Calibrators.}
\label{calibrators}
\begin{center}
\begin{tabular}{@{}cccccc}
Calibrator\(^{\rm{a}}\) & $\alpha$ (B1950)\(^{\rm{a}}\) & $\delta$ (B1950)\(^{\rm{a}}\) & $S_{843}$ Nominal\(^{\rm{a}}\) & $S_{843}$ Measured\(^{\rm{b}}\) & $\sigma_{\rm{rms}}$\(^{\rm{c}}\) \\
Name   & $h$ $m$ $s$ &  $^\circ$ $'$ $\arcsec$ & (Jy) & (Jy) & per cent of Measured \\
\hline
0252--712 & 02 52 26.63 & $-71$ 16 47.3 & 9.21 & $9.13\pm0.01$ & 2.1 \\
0409--752 & 04 09 58.45 & $-75$ 15 05.7 & 19.80 & $20.21\pm0.03$ & 2.4 \\
0420--625 & 04 20 18.61 & $-62$ 30 40.9 & 5.62 & $5.49\pm0.02$ & 2.8 \\
1814--519 & 18 14 07.92 & $-51$ 59 20.0 & 6.51 & $6.55\pm0.02$ & 3.7 \\
1814--637 & 18 14 45.94 & $-63$ 47 03.1 & 20.22 & $20.04\pm0.05$ & 3.6 \\
1827--360 & 18 27 36.86 & $-36$ 04 38.1 & 13.86 & $13.72\pm0.06$ & 3.6\\
2323--407 & 23 23 51.69 & $-40$ 43 48.8 & 5.21  & $5.16\pm0.03$ & 3.8 \\
\end{tabular}
\end{center}

\medskip
NOTES:\\
\(^{\rm{a}}\)The source names, positions
  and values of
  nominal flux density are taken from~\citet{wilson94}.\\
\(^{\rm{b}}\)The values of measured 
  flux density have been corrected for effects described
  in~\citet{gaensler2000}
  and are average measured values for the 12-year period.\\
\(^{\rm{c}}\)The rms
  in flux density measurements 
  as a percentage of measured flux density 
  from observations, used to define the 
  calibration uncertainty of the MOST.\\
\end{minipage}
\end{table*}

\subsection{Flux Density}

The uncertainties in the fitted peak brightness in SUMSS images are calculated
as the quadratic sum of the MOST internal flux density
calibration uncertainty and the local
noise uncertainty. The local noise uncertainty is calculated using 
\begin{equation}
\sigma_{A_{843}}^2=\frac{A_{843}^2}{\rho^2},
\label{peakerr}
\end{equation}
where $\alpha_M=\alpha_m=3/2$ was used in the calculation of $\rho^2$ from
Equation~\ref{sn}~\citep{condon98}.

To calculate the flux density calibration uncertainty of the
MOST we used results from a
detailed analysis of the Molonglo calibrators \citep{gaensler2000}. 
Before and after every 12-hour 
observation the MOST measures the flux densities of
$\sim$5 compact sources, chosen from a list of 55 calibrators. Observations
of the calibrators in the period 1984 to 1996 have been extracted from the
MOST archive and used to
determine the calibration uncertainty of the MOST.
\citet{gaensler2000} examined the variability of the calibrators in this
period and found that 19 of these showed no variability in this time.
We have
chosen 7 of the non-variable calibrators with flux densities
$>5$~Jy which were
observed at a meridian distance $|{\rm MD}|<30^{\circ}$ to ensure that
errors resulting from fan-beam 
confusion and uncertainty in the meridian distance
gain curve at higher MD were minimised. 
Table~\ref{calibrators} shows the results of this analysis.

The average scatter in flux density measurements \(\epsilon_{A_{843}}\) 
is around 3 per cent. We
have adopted this value as the internal 
calibration uncertainty of MOST peak 
brightness 
measurements. Peak brightness uncertainties in the catalogue are obtained
by adding $\epsilon_{A_{843}}A_{843}$ and
\(\sigma_{A_{843}}\) in quadrature. 

The integrated flux density of each source is calculated from the
parameters of the gaussian fit and depends on
whether or not the source is significantly resolved. We use the same
equations to derive the total flux density
as those described for
NVSS \citep{condon98}. The fitting
uncertainties in integrated flux density are the same as those
quoted in C97.

The quoted flux density uncertainties do not take account of the errors which
arise from fitting an extended gaussian to a source with complex
structure. Extended sources are often
fitted poorly by an elliptical gaussian model and this can lead to
unreliable estimates of the true integrated flux density. Users should
therefore note that for some extended
sources the quoted flux density and source sizes
will be incorrect by more than the quoted uncertainties.

The flux density scale at 843\,MHz was determined partly by absolute
measurements and partly by interpolation between measurements at 408\,MHz
(Molonglo) and 2700\,MHz (Parkes) \citep{hunstead91}. To check the accuracy of the
total flux densities quoted in the SUMSS catalogue we have examined the distribution
of spectral indices between the SUMSS catalogue at 843\,MHz and the NVSS catalogue at
1.4\,GHz. Flux densities in the NVSS catalogue are known to be accurate to 2
per cent, given
the same caveats associated with {\sc vsad} explained in this paper \citep{condon98}.
%Ideally it would be possible to examine the flux calibration of the SUMSS
%survey by comparison with other observations at 843MHz made with different 
%instruments. Unfortunately it is difficult to show this for the
%SUMSS survey as there is no other large survey at the same frequency which
%can be used for comparison.
Figure~\ref{sindex} shows the distribution of spectral index ($\alpha$)\footnote{In this
  paper we define spectral index $\alpha$ to be $S_{\nu}\propto\nu^{\alpha}$}
  between SUMSS and NVSS in three flux density bins. This
  was determined by crossmatching all sources in the SUMSS-NVSS overlap
  region ($-40^{\circ}\leq\delta\leq-30^{\circ}$) with 
  position offset no greater than 30$\arcsec$ and only a single match
  within 100$''$. This
  resulted in 7643 matches. The overall median spectral index is
  $-0.83$, which is consistent with previous determinations of
  spectral index at frequencies below 1.4\,GHz \citep{oort88,hunstead91,debreuck2000}. A
  tail of flat spectrum $(\alpha \sim 0)$ sources (probably QSOs) can be seen in the top
  panel. The FWHM of the distributions increases from 0.7 in the upper
  panel to 1.2 in the lower panel
  reflecting the increasing uncertainty in flux density for 
  $S_{843}<20$\,mJy. The steep and flat spectrum tails ($\alpha<-2.0$ and
  $\alpha>1.0$ respectively) in the
  bottom panel were checked visually and found to be due to fitting errors and
  erroneous flux densities in both NVSS
  and SUMSS.

\begin{figure}
\includegraphics[width=\linewidth,height=10cm]{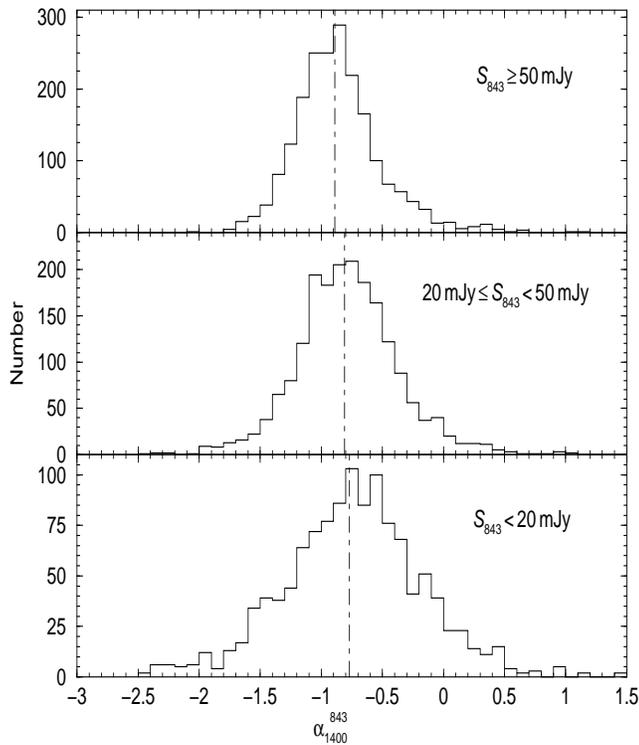}
\caption{The spectral index distribution in the NVSS-SUMSS overlap
  region binned into three flux density ranges. A dotted line showing the
  median spectral index in each flux density bin is shown. The median
  spectral index increases from $-$0.89 at $S_{843}\geq50$ mJy to
  $-$0.77 at $S_{843}<20$ mJy. A tail of flat spectrum sources can be
  seen for $S_{843}\geq50$ mJy. 
  The scatter in the spectral index increases with decreasing flux density.}
\label{sindex}
\end{figure}

\section{Catalogue Format}

%\begin{twocolumn}
%\begin{table*}
%\vbox to220mm{\vfil Landscape table to go here.
%  \caption{}
%\vfil}
%\label{sumsscat}
%\end{table*}

Table~3 shows the format of the SUMSS catalogue. The catalogue
will be available as a large text file accessible via the web. A
short description of each of the columns of the catalogue follows.

\medskip

\noindent  \textit{Columns} (1) \& (2): The right ascension ($\alpha$) and declination
  ($\delta$) of the source in J2000 coordinates.
  
\noindent  \textit{Column} (3): The uncertainty in Right Ascension in seconds of arc,
  calculated 
    from the quadratic sum of the MOST Right Ascension calibration uncertainty $\left(1\farcs1\right)$ and
    equation~\ref{raerr}.
  
\noindent  \textit{Column} (4): The uncertainty in Declination in seconds of arc, calculated
    from the quadratic sum of the MOST Declination calibration uncertainty
    $\left(1\farcs6\right)$ and equation~\ref{decerr}.

\noindent  \textit{Column} (5): The peak brightness 
  in units of mJy beam$^{-1}$ and its associated uncertainty
    calculated from the quadratic sum of equation~\ref{peakerr} and the MOST flux
  density calibration uncertainty of 3 per cent.
 
\noindent  \textit{Column} (6): The total flux density
  in units of mJy and its associated uncertainty, calculated
    from the equations described in C97. 
  
\noindent  \textit{Columns} (7) \& (8): The fitted major \& minor axes of the
    source in arcseconds.
 
\noindent  \textit{Column} (9): The fitted major axis position angle of the source in 
   degrees east of north. Most
    unresolved sources would have PA values close to 0$^{\circ}$ or
    180$^{\circ}$ since the MOST elliptical beam has PA=0.

\noindent  \textit{Column} (10): If the fitted major axis size exceeds the
   beam size by more than $2.33\sigma(\theta_{\rm{M}})$, 
   the major axis size after 
   deconvolution from the MOST beam is given in arcseconds.

\noindent   \textit{Column} (11): If the major axis is resolved the minor axis is
   subsequently checked using the same criterion. If
   the minor axis is found to be resolved the deconvolved minor axis size
   is given in arcseconds.

\noindent   \textit{Column} (12): If the major axis is resolved, its deconvolved
   position angle in degrees east from north is given.
  
\noindent   \textit{Column} (13): The name of the mosaic in which the source appears. The
    original 
    mosaics are available online at
    http://www.astrop.physics.usyd.edu.au/mosaics. In the case of
    duplicate matches the mosaic name quoted is that used for the fit which is
    chosen to be included in the catalogue. 
  
\noindent   \textit{Column} (14): The number of mosaics in which the source appears. This is
    included to let the user know when the source appears in more than one
    image. The source parameters which appear in the catalogue are those
    for the most reliable fit.
  
\noindent   \textit{Columns} (15) \& (16): The X \& Y pixel positions of the source on the
   quoted mosaic.

\medskip

% mnland.tex, landscape material for sample pages
% v1.1 released 18th July 1994
% v1.0 released 28th January 1994

%\documentstyle[landscape,amssym,rotating]{mn}

%\ifoldfss    
%  \newcommand{\rmn}[1] {{\rm #1}}
%\else
%  \newcommand{\rmn}[1] {\mathrm{#1}}
%\fi

%\pagestyle{empty}

%\begin{document}

% only use the following if your dvi driver understands `landscape'
%\hbox{\special{landscape}}
%\clearpage

\setcounter{table}{3}
\begin{onecolumn}
\begin{sidewaystable}
%\vspace*{-1.5cm}
%\hspace{2cm}
\begin{minipage}{225mm}
{\bf Table 3.} The First Page of the SUMSS Catalogue.\\
\label{sumsscattable}
\begin{tabular}{@{}ccccrrrrrrrrrrccrr}
\hline
(1) & (2) & (3) & (4) & \multicolumn{1}{c}{(5)} & \multicolumn{1}{c}{(6)} & \multicolumn{1}{c}{(7)} & \multicolumn{1}{c}{(8)} & \multicolumn{1}{c}{(9)} & \multicolumn{1}{c}{(10)} & \multicolumn{1}{c}{(11)} & \multicolumn{1}{c}{(12)} &
\multicolumn{1}{c}{(13)} & \multicolumn{1}{c}{(14)} & \multicolumn{1}{c}{(15)} &
\multicolumn{1}{c}{(16)} & \multicolumn{1}{c}{(17)} & \multicolumn{1}{c}{(18)} \\
\(\alpha\) (J2000) & \(\delta\) (J2000) & \(\Delta\alpha\) & \(\Delta\delta\) & \multicolumn{1}{c}{\(A{_{843}^{\rm{a}}}\)} &
\(\sigma{_A}\) & \multicolumn{1}{c}{\(S{_{843}^{\rm{b}}}\)} & \(\sigma{_S}\) &
\multicolumn{1}{c}{\(\theta{_{\rm{M}}}^{\rm{c}}\)} &
\multicolumn{1}{c}{\(\theta{_{\rm{m}}}^{\rm{c}}\)} &
\multicolumn{1}{c}{\(\rm{PA}{_{\rm{F}}}^{\rm{c,d}}\)} & 
\multicolumn{1}{c}{\(\phi{_{\rm{M}}}^{\rm{e}}\)} & \multicolumn{1}{c}{\(\phi{_{\rm{m}}}^{\rm{e}}\)}
& \multicolumn{1}{c}{\(\rm{PA}{_{\rm{S}}}^{\rm{e,d}}\)} & Mosaic\(^{\rm{f}}\) & \#\(^{\rm{f}}\) & X-Pixel & Y-Pixel \\
\(h \,\,\, m \,\,\, s\) & \(^{\circ} \,\,\, \arcmin \,\,\, \arcsec\) & \arcsec & \arcsec &
\multicolumn{2}{c}{mJy beam\(^{-1}\)} & \multicolumn{2}{c}{mJy} &
\multicolumn{1}{c}{\arcsec} & \multicolumn{1}{c}{\arcsec}
& \multicolumn{1}{c}{\(^\circ\)} & \multicolumn{1}{c}{\arcsec} &
\multicolumn{1}{c}{\arcsec} & \multicolumn{1}{c}{\(^\circ\)} & & & & \\
\hline

00 00 00.00 & $-$31 09 52.24 & 4.4 & 4.6 & 12.1 & 1.4 & 13.8 & 1.6 & 84.9 & 58.4 & 41.3 & 0.0 & 0.0 & \texttt{---} & J0000M32 &1 & 705.0 & 518.9 \\
00 00 02.31 & $-$37 08 00.38 & 2.3 & 3.4 & 13.2 & 1.2 & 13.3 & 1.2 & 76.6 & 45.3 &177.2 & 0.0 & 0.0 & \texttt{---} & J0000M36 &1 & 702.5 & 197.0 \\
00 00 03.28 & $-$75 01 02.93 & 1.8 & 2.1 & 29.4 & 1.5 & 33.0 & 1.7 & 58.6 & 45.0 & 5.4 & 35.8 & 0.0 & 5.9 & J0000M76 &1 & 703.8 & 996.0 \\
00 00 03.29 & $-$66 05 43.51 & 1.7 & 1.8 & 25.6 & 1.0 & 25.7 & 1.0 & 50.1 & 45.0 & 90.7 & 0.0 & 0.0 & \texttt{---} & J0000M64 &2 & 703.2 & 17.8 \\
00 00 04.14 & $-$64 19 06.06 & 3.5 & 3.9 & 7.4 & 0.9 & 7.4 & 0.9 & 50.1 & 45.1 &173.1 & 0.0 & 0.0 & \texttt{---} & J0000M64 &1 & 702.6 & 540.4 \\
00 00 04.61 & $-$75 12 43.42 & 4.3 & 4.8 & 9.7 & 1.4 & 11.4 & 1.6 & 57.4 & 50.1 & 16.1 & 0.0 & 0.0 & \texttt{---} & J0000M76 &1 & 703.4 & 934.2 \\
00 00 06.14 & $-$29 55 10.49 & 4.5 & 7.6 & 11.8 & 2.1 & 13.1 & 2.3 & 93.0 & 51.0 &169.4 & 0.0 & 0.0 & \texttt{---} & J0000M32 &1 & 697.7 & 734.7 \\
00 00 06.39 & $-$34 19 06.49 & 4.0 & 5.8 & 10.4 & 1.6 & 10.9 & 1.7 & 76.6 & 49.5 & 11.6 & 0.0 & 0.0 & \texttt{---} & J0000M36 &1 & 697.8 & 738.4 \\
00 00 06.39 & $-$63 11 52.08 & 1.8 & 1.8 & 34.0 & 1.3 & 43.8 & 1.7 & 75.9 & 49.5 & 67.2 & 60.6 & 0.0 & 69.9 & J0000M64 &1 & 701.1 & 870.0 \\
00 00 06.43 & $-$69 25 35.54 & 1.5 & 1.7 & 118.9 & 3.6 & 118.9 & 3.6 & 48.6 & 45.0 & 91.6 & 0.0 & 0.0 & \texttt{---} & J0000M68 &1 & 701.9 & 221.2 \\
00 00 07.78 & $-$70 21 10.15 & 1.9 & 2.1 & 18.2 & 0.9 & 18.2 & 0.9 & 47.3 & 45.3 & 17.9 & 0.0 & 0.0 & \texttt{---} & J0000M72 &1 & 701.4 &1182.6 \\
00 00 07.96 & $-$72 36 43.24 & 2.2 & 2.3 & 14.2 & 0.9 & 14.5 & 0.9 & 48.0 & 46.0 & 38.7 & 0.0 & 0.0 & \texttt{---} & J0000M72 &1 & 701.8 & 479.5 \\
00 00 08.22 & $-$37 38 19.61 & 3.1 & 3.3 & 18.1 & 1.5 & 20.8 & 1.7 & 76.6 & 59.9 & 41.6 & 0.0 & 0.0 & \texttt{---} & J0000M36 &1 & 696.1 & 99.8 \\
00 00 08.89 & $-$71 00 19.76 & 2.3 & 2.5 & 13.2 & 0.9 & 13.3 & 0.9 & 47.3 & 45.6 &151.2 & 0.0 & 0.0 & \texttt{---} & J0000M72 &1 & 701.1 & 979.5 \\
00 00 09.90 & $-$31 33 30.53 & 1.6 & 2.1 & 44.7 & 2.0 & 44.7 & 2.0 & 84.9 & 45.1 & 0.9 & 0.0 & 0.0 & \texttt{---} & J0000M32 &1 & 693.5 & 450.6 \\
00 00 10.70 & $-$63 04 14.09 & 2.8 & 2.9 & 11.3 & 1.0 & 13.0 & 1.1 & 62.4 & 47.6 & 42.1 & 0.0 & 0.0 & \texttt{---} & J0000M64 &1 & 698.4 & 907.4 \\
00 00 10.86 & $-$72 12 53.71 & 1.7 & 1.9 & 25.0 & 1.1 & 25.2 & 1.1 & 47.3 & 45.5 &119.1 & 0.0 & 0.0 & \texttt{---} & J0000M72 &1 & 700.5 & 603.1 \\
00 00 11.43 & $-$85 39 20.09 & 1.5 & 1.7 & 101.6 & 3.1 & 102.9 & 3.2 & 46.3 & 45.2 & 65.9 & 0.0 & 0.0 & \texttt{---} & J0000M84 &3 & 703.8 & 162.2 \\
00 00 11.82 & $-$82 47 32.39 & 1.7 & 1.8 & 74.4 & 2.7 & 101.0 & 3.7 & 69.0 & 54.4 & 66.9 & 52.3 & 30.2 & 67.2 & J0000M84 &1 & 703.0 &1094.0 \\
00 00 11.93 & $-$66 30 45.94 & 1.5 & 1.7 & 93.5 & 2.9 & 98.2 & 3.1 & 53.0 & 45.4 & 5.7 & 21.4 & 0.0 & 10.1 & J0000M68 &1 & 698.5 &1105.2 \\
00 00 13.15 & $-$63 34 57.29 & 3.6 & 3.5 & 10.3 & 1.0 & 13.4 & 1.4 & 79.8 & 48.2 & 48.8 & 0.0 & 0.0 & \texttt{---} & J0000M64 &1 & 697.0 & 756.8 \\
00 00 13.17 & $-$72 59 54.71 & 1.8 & 1.9 & 28.1 & 1.2 & 29.9 & 1.3 & 51.2 & 47.2 &103.6 & 0.0 & 0.0 & \texttt{---} & J0000M72 &1 & 699.7 & 359.2 \\
00 00 13.29 & $-$35 55 20.96 & 3.8 & 3.1 & 12.2 & 1.3 & 12.6 & 1.3 & 76.6 & 48.3 & 63.2 & 0.0 & 0.0 & \texttt{---} & J0000M36 &1 & 690.3 & 429.9 \\
00 00 14.02 & $-$34 10 00.23 & 1.7 & 2.0 & 65.4 & 2.5 & 77.0 & 3.0 & 76.6 & 62.4 &150.2 & 0.0 & 0.0 & \texttt{---} & J0000M36 &1 & 689.2 & 767.6 \\
00 00 15.68 & $-$76 56 30.59 & 1.5 & 1.7 & 84.5 & 2.7 & 86.3 & 2.8 & 48.3 & 45.1 & 11.9 & 0.0 & 0.0 & \texttt{---} & J0000M76 &1 & 700.2 & 384.9 \\
00 00 15.77 & $-$33 12 21.60 & 3.2 & 2.3 & 17.0 & 1.4 & 17.1 & 1.4 & 84.9 & 45.7 & 78.0 & 0.0 & 0.0 & \texttt{---} & J0000M32 &1 & 687.0 & 164.9 \\
00 00 15.86 & $-$70 49 28.88 & 2.8 & 3.2 & 10.0 & 1.0 & 10.5 & 1.0 & 52.0 & 45.2 & 14.1 & 0.0 & 0.0 & \texttt{---} & J0000M72 &1 & 697.9 &1035.8 \\
00 00 17.30 & $-$82 40 59.16 & 2.2 & 2.3 & 23.8 & 1.5 & 23.8 & 1.5 & 45.2 & 45.0 & 65.8 & 0.0 & 0.0 & \texttt{---} & J0000M84 &1 & 702.0 &1129.6 \\
00 00 17.37 & $-$37 28 25.07 & 2.8 & 4.2 & 11.2 & 1.3 & 11.3 & 1.3 & 76.6 & 45.7 & 4.3 & 0.0 & 0.0 & \texttt{---} & J0000M36 &1 & 686.2 & 131.6 \\
00 00 17.76 & $-$34 10 40.26 & 1.6 & 1.7 & 126.4 & 4.1 & 132.4 & 4.3 & 76.6 & 49.4 &116.7 & 0.0 & 0.0 & \texttt{---} & J0000M36 &1 & 685.0 & 765.5 \\
00 00 17.84 & $-$35 18 07.16 & 4.0 & 3.7 & 14.5 & 1.5 & 16.3 & 1.7 & 76.6 & 57.1 &122.0 & 0.0 & 0.0 & \texttt{---} & J0000M36 &1 & 685.1 & 549.3 \\
00 00 17.94 & $-$37 21 00.68 & 3.0 & 3.5 & 16.7 & 1.4 & 19.3 & 1.6 & 76.6 & 59.9 &154.1 & 0.0 & 0.0 & \texttt{---} & J0000M36 &1 & 685.6 & 155.3 \\

\hline
\end{tabular}

\flushleft
NOTES:\\
\(^{\rm{a}}\) The peak brightness of the gaussian fit in units of mJy beam\(^{-1}\). This value may be
in error by more than the quoted error for extended sources.\\
\(^{\rm{b}}\) The total flux density of the gaussian fit in units of
mJy. \(S = A\) for point sources.\\
\(^{\rm{c}}\) The widths and position angle of the fitted gaussian. The fit
is constrained so that \(\theta_{\rm{m}}\geq45''\) (the beam minor axis width).\\
\(^{\rm{d}}\) The position angle of the major axis is measured in degrees East from North.\\
\(^{\rm{e}}\) The deconvolved widths and position angle of the source. A
value is given only if the fitted axis exceeds the beam by more than 2.33\(\sigma_{\theta}\)\\
\(^{\rm{f}}\) The name of the mosaic the quoted source can be found
in. If the number in the next column
is greater than 1 it can also be found in neighbouring mosaics.\\

\end{minipage}

\end{sidewaystable}

\end{onecolumn}

%\end{document}

\begin{twocolumn}

Please refer to SUMSS catalogue sources by their full IAU
designations \citep{lortet94}. These are of the form SUMSS {\it JHHMMSS$-$DDMMSS} where
SUMSS is the survey acronym, {\it J} specifies J2000.0 coordinate equinox,
{\it HHMMSS} are the hours, minutes and truncated seconds of right
ascension, $-$ is the sign of declination and {\it DDMMSS} are the degrees,
minutes and truncated seconds of declination. For example the SUMSS source
in Table 3 at J2000.0 coordinates $\alpha=00^h00^m08\fs89, \,\,\,
\delta=-71^{\circ}00'19\farcs76$ is called SUMSS J000008$-$710019.

\section{Analysis}
\label{analysis}

The following section contains results of our analysis of the catalogue.
Our goal in producing the SUMSS catalogue has been to create a source list
which is as reliable as possible. 
This implies that all the sources which appear should be genuine. In this
section we also examine the uniformity of the catalogue. An adequate
determination of 843\,MHz source counts and the two-point angular
correlation function will require a catalogue with uniform source density 
\citep{blake02}. Finally the SUMSS catalogue is compared
with catalogues at other frequencies as an independent check of the accuracy
of quoted source characteristics.

\subsection{Reliability and Completeness}

The rms noise level in SUMSS mosaics is not uniform across the survey. It 
can increase substantially close to bright sources and changes strongly
with declination (see Figure~\ref{rmsdec}). In surveys with uniform
noise levels a 5$\sigma$ limiting amplitude is applied to catalogues
made from them \citep{murdoch73}. As the noise is not
uniform in SUMSS the limiting peak amplitude varies from 4$\sigma$ to 6$\sigma$
depending on the position of each source in the survey. 

To estimate how the SUMSS catalogue reliability varies with amplitude
we have searched 5 northern mosaics to 2$\sigma$ and crossmatched
all fitted sources with the NVSS. The NVSS is believed to be better than 
90 per cent reliable at flux densities above 5\,mJy at 1.4\,GHz \citep{condon98} 
and so should be a good
independent check of the reliability of SUMSS. We define the reliability at
amplitude $A_{843}$ as the number of SUMSS sources with an NVSS
counterpart within 50$''$ divided by the total number of SUMSS sources. 
There is about a 3 per cent chance of an NVSS position being within 
50$''$ of a random SUMSS noise peak.

Figure~\ref{reliability} shows a plot of the reliability of SUMSS as a
function of peak flux density. At the 4$\sigma$
level, the catalogue is still around 80 per cent reliable and this
increases to 90 per cent at around 4.5$\sigma$. The SUMSS catalogue
is therefore still reliable below 5$\sigma$, predominantly because the decision
tree is trained to remove spurious responses.

Fitting errors and confusion in the MOST beam
can affect the completeness of the catalogue at fainter flux densities. 
The decision tree is 
geared towards removing poorly fitted sources at lower flux density and
this may
also affect the completeness.
We have attempted to determine the completeness of the SUMSS catalogue by
running simulations. We placed 1000 artificial point sources of varying
flux density into a selection of mosaics and performed the normal
cataloguing procedures on them. 

Figure~\ref{completeness} shows the fraction of artificial point sources
which were catalogued vs. flux density. In fitting the 1000 point sources
no particular biases were evident. The shape of the distributions is quite different for
northern and southern mosaics. For southern mosaics the plot shows that the
SUMSS catalogue is
$\sim60$ per cent complete at the limiting flux density
of 6\,mJy and this increases to 100 per cent at
8\,mJy. For northern mosaics the catalogue is $\sim40$ per cent complete at
10\,mJy and 100 per cent complete at 18\,mJy. The more extended distribution in the
northern catalogue is because the mosaics here are noisier and the beam is
larger, resulting in greater confusion.
The decision tree has had no significant detrimental effect on the completeness for
point sources. This is because artefacts generally mimic extended
sources and the decision tree tends to remove extended sources at lower
flux density rather than point sources.

\begin{figure}
\includegraphics[width=\linewidth]{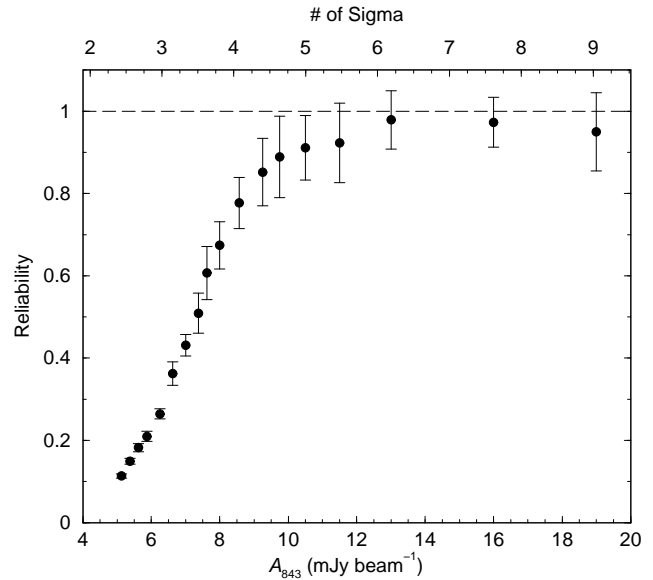}
\caption{The fraction of SUMSS sources detected in the NVSS catalogue
  vs. peak flux density. At the catalogue limit of 10 mJy beam$^{-1}$ about
  90 per cent of SUMSS sources are detected in NVSS. Sigma has been
  calculated as the median local rms noise of all sources in this plot 
  (1$\sigma$=2.1\,mJy\,beam$^{-1}$). 
  At these flux
  densities the NVSS is also about 90 per cent reliable.}
\label{reliability}
\end{figure}

\begin{figure}
\includegraphics[width=\linewidth]{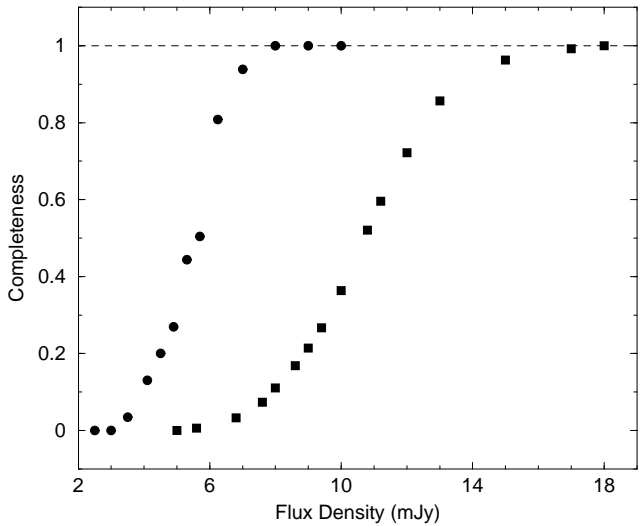}
\caption{The fraction of artificial point sources which are 
  catalogued vs. Flux Density. The circles are for southern mosaics
  $(\delta\leq-50^{\circ})$
  and the squares are for northern ones $(\delta>-50^{\circ})$. 
  A dashed line at 100 per cent is
  shown.
  At 6mJy (the limit of the southern
  catalogue)
  the completeness for southern mosaics is around 60 per cent and this rises
  rapidly to 100 per cent at 8\,mJy. At 10\,mJy the completeness for the northern
  catalogue
  is about 40 per cent. The northern catalogue is 100 per cent complete above about 18\,mJy.}
\label{completeness}
\end{figure}

\subsection{Source Density}

\begin{figure*}
\includegraphics[width=8cm]{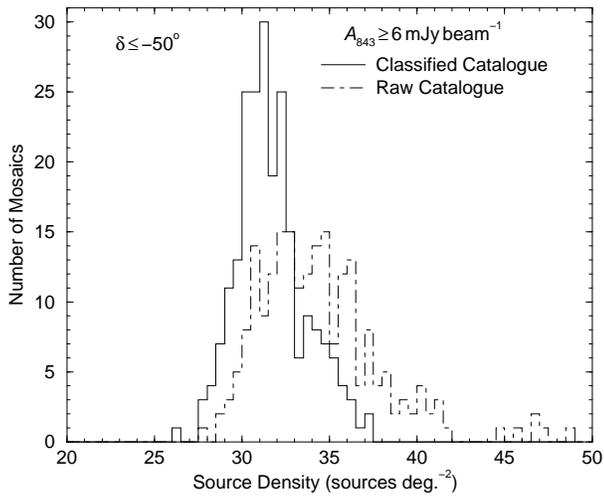}%
\hspace{1cm}
\includegraphics[width=8cm]{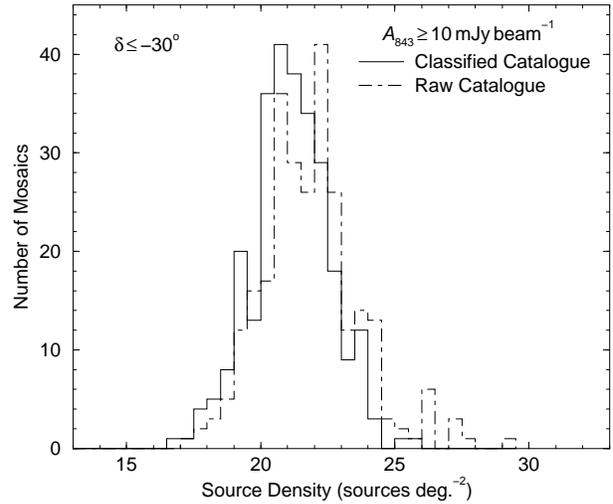}
\caption{Histograms showing the source density of the SUMSS catalogue
  before and after the application of the decision tree. The source density
  was calculated for each mosaic and a random
    selection of non-overlapping mosaics
  was chosen so as to keep the data independent. The plot on the left is
  for southern sources with $A_{843}\geq6$\,mJy\,beam$^{-1}$. 
  Northern and southern sources are grouped in the plot on the right,
  which shows the source density for sources with
  $A_{843}\geq10$\,mJy\,beam$^{-1}$.}
\label{density}
\end{figure*}

Table 4 of Paper~I~\citep{bock99} shows integral source counts at 843MHz taken from Large
(1990) and we are now able to compare the source counts in that table with
those derived in the SUMSS survey. The source density has been determined
from a selection of non-overlapping mosaics in the catalogue and the results of this analysis
are shown in Figure~\ref{density}. To examine the effect of the decision
tree on the source density, we have plotted the source density both before
and after its application. The source density derived from the raw catalogue will contain
artefacts as well as real sources while the classified catalogue should
only have real sources, although some real sources may have been erroneously removed by the
decision tree.

For $S_{843}\geq6$\,mJy Large (1990) estimated an 843\,MHz source density of
$31\pm3$ sources~deg$^{-2}$. 
Figure~\ref{density}~(left) shows the source density in
southern mosaics limited to peak brightness
$A_{843}\geq6\,\rm{mJy}\,\rm{beam}^{-1}$. The distribution
before application of the decision tree has an average of
$36.3\pm0.6$ sources deg$^{-2}$ and, after classification, 
$31.6\pm0.1$ deg$^{-2}$. Outliers in the distribution are
attributed to mosaics with strong sources which produce many artefacts.  
In the raw catalogue these artefacts are fitted as sources, producing
an increased source density for that mosaic. In the classified catalogue, the
source density is not uniform because the local rms noise close to strong
sources is higher than in other regions.

%limited dynamic range of MOST has meant that the density of
%`real' sources left in the catalogue by the decision tree is less than
%the average source density.

The effect of the decision tree is less pronounced at
$A_{843}\geq10$\,mJy\,beam$^{-1}$. Large (1990) determined a source density of $21\pm2$\,deg$^{-2}$. The plot on the right in Figure~\ref{density} shows
the distribution of source densities in the same set of southern 
mosaics with the 
addition of
non-overlapping northern
mosaics with $A_{843}\geq10\,\rm{mJy}\,\rm{beam}^{-1}$. For
$A_{843}\geq10$\,mJy\,beam$^{-1}$ 
the distribution is very similar for both the
raw and classified catalogues and indicates that our decision tree is
predominantly affecting sources of peak brightness 
less than 10\,mJy\,beam$^{-1}$  (see Figure~\ref{lognlogs}). The
average source density 
for the raw catalogue is $22.1\pm0.2$\,deg$^{-2}$ 
and the average for the
classified catalogue is $21.1\pm0.1$\,deg$^{-2}$.

Our results are in good agreement with
those derived by Large (1990) from a much smaller survey area.
One important aspect of the distribution of source density in
the classified catalogue is that there is less scatter between mosaics
after the application of the decision tree. This indicates that the decision
tree is making the catalogue more uniform.

\begin{figure}
\includegraphics[width=\linewidth]{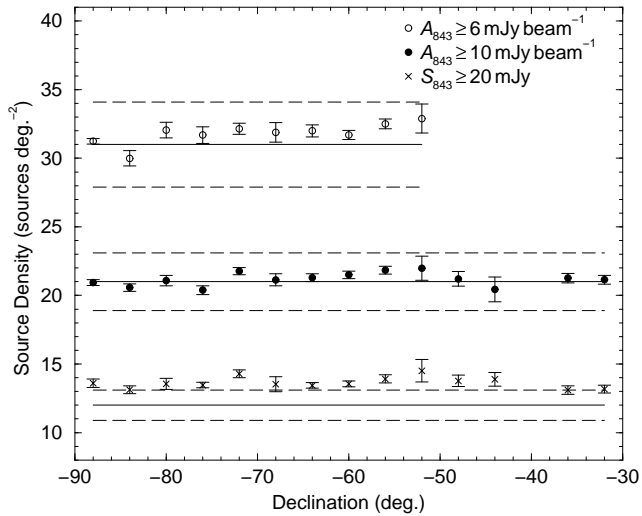}
\caption{The variation in source density with declination for different
  flux density limits shows little
  scatter across the survey. This is primarily because the
  telescope does not change its configuration at different
  declinations. The lines are the values of source count values from Large
  (1990) with their associated 10 per cent uncertainty. A peak brightness
  cutoff was applied at both 10 and 6 mJy beam$^{-1}$ and a total flux
  density cutoff was applied at 20 mJy, where we expect the catalogue to be
  complete.}
\label{counts}
\end{figure}

Figure~\ref{counts} shows the variation of source density in the catalogue
with declination. This has been determined for three
different flux density cutoffs 
($A_{843}\geq$6\,mJy\,beam$^{-1}$, $A_{843}\geq$10\,mJy\,beam$^{-1}$ \& $S_{843}\geq$20\,mJy). 
The lines in the plot are the values of source
density from Large 
(1990) for the three flux density cutoffs.
We expect the SUMSS catalogue to be 100 per cent complete above
20\,mJy so we directly compare the counts for integrated flux density with
those quoted by Large (1990). We believe the source counts quoted by Large
(1990) may have been underestimated due to the small area of sky used
(about 28\,deg$^2$).

%Our results are systematically higher than
%the Large (1990) value. We believe this may be due to the small area of sky
%used by Large (1990) (about 28\,deg$^2$), 
%which could lead to an underestimate of the
%source counts at higher flux densities. 

Figure~\ref{counts} is similar to that made for the NVSS survey by
\citet{blake02} but does not show the same scatter evident in their
results. \citet{blake02} attributed their results to the change from the
DnC to the D configuration of the VLA for the NVSS survey at different
declinations. No such 
change takes place on the MOST, thereby resulting in a survey which is
more uniform with declination,
especially at flux densities greater than 10mJy.

%There is a slight increase in the source
%density with declination at all three flux levels. We have attributed this
%to confusion in the enlarged MOST beam at northern declinations. Two sources
%with flux density below the 6mJy limit of the survey will both contribute
%to a point source i
%%
%the larger beam at northern declinations thereby increasing the source
%
%density in the
%
%catalogue. 

\subsection[]{SUMSS--MRC Crossmatch}

\begin{table}
\begin{minipage}{80mm}
\caption{MRC sources missing from the SUMSS catalogue}
\label{sumssmrc}
\begin{center}
\begin{tabular}{@{}lccc}
\hline
MRC Name & $\alpha$ (J2000) & $\delta$ (J2000) & $S_{408}$ \\
 & $h$ $m$ $s$ & $^\circ$ $'$ $''$ & Jy \\
\hline
MRC\,B0536$-$692$^1$ & 05 36 20.7 & $-$69 12 15 & 1.03 \\
MRC\,B0540$-$697$^1$ & 05 39 58.5 & $-$69 45 00 & 2.22 \\
MRC\,B1737$-$602 & 17 42 02.9 & $-$60 15 54 & 0.80 \\
MRC\,B1754$-$577$^2$ & 17 59 05.2 & $-$57 42 20 & 1.21 \\
MRC\,B1817$-$632$^2$ & 18 22 16.0 & $-$63 10 41 & 0.90 \\
MRC\,B2220$-$700 & 22 24 39.6 & $-$69 47 42 & 0.73 \\
\hline
\end{tabular}
\end{center}
\medskip
NOTES:\\
$^1$ \hbox{H\,{\sc ii}} regions located in the LMC.\\
$^2$ Spurious north-south sidelobes of the Mills Cross.
\end{minipage}
\end{table}

As a separate check of the catalogue completeness for brighter sources, the
SUMSS catalogue was crossmatched with the Molonglo Reference Catalogue
(MRC; \citealt{mrc81,mrc91}). The MRC was made from
$2\farcm62\times2\farcm86\,\rm{sec}\left(\delta+35\fdg5\right)$ resolution observations at 408 MHz
using the Mills Cross Radio Telescope \citep{cross63}, the previous incarnation of
the MOST. 
The MRC is complete to $S_{408}=1\,\rm{Jy}$ at 408~MHz and has a
(non-uniform)
limiting flux
density of $S_{408}=0.7\,\rm{Jy}$. Given a spectral index of $\alpha=-0.8$ we
expect the faintest MRC sources to appear in the SUMSS catalogue at around
$S_{843}=400$~mJy and  certainly no fainter than $S_{843}=150$\,mJy. This
implies that all MRC sources should appear in the SUMSS catalogue.

There are 1670 MRC sources in the area covered by the SUMSS
catalogue. Of these only 46 were found not to have a match within
60$''$ in SUMSS. Upon closer inspection it was revealed that
many of these initial non-detections were actually sources which were
resolved into doubles in SUMSS; for these sources there are two entries in
the SUMSS catalogue around 100$''$ from the single source in the MRC. 
Of the sources which are detected, we determine a median spectral index
between 408\,MHz and 843\,MHz of $\alpha=-0.95$.

Two of the missing MRC sources (Centaurus A \& NGC 5090) were found to be
fitted quite badly by {\sc vsad}. Fits to these sources were modified.
They have not been assigned a peak amplitude or source size, 
only a position and total
flux density. The total flux density 
of these sources has been determined by summing the pixels inside a 
hand defined source area using the {\sc cgcurs} routine in {\sc miriad}. 
We may add other complex sources by hand in
future releases of the catalogue.

The
MRC sources not detected in SUMSS are tabulated in
Table~\ref{sumssmrc}. Some of the sources are \hbox{H\,{\sc ii}} regions in 
the Large Magellanic Cloud which, although
point sources in the MRC, appear quite complex in the smaller MOST
beam. Elliptical gaussians fitted to these sources were automatically
removed by the decision tree and have been left out of the catalogue. 

The
four sources away from the Clouds are particularly interesting. The positions
of these sources have been inspected in the original SUMSS images and are
not detected at even a 3$\sigma$ limit. 
Two of the sources, MRC\,B1817--632 and MRC\,B1754--577 were observed with
the Australia Telescope Compact Array 
at 1.4\,GHz and no source was detected to a $3\sigma$ limit of 0.5\,mJy
at both MRC positions. 
All of the original 408\,MHz data obtained with the Mills
Cross telescope are currently being reprocessed by D. Crawford (private
communication) and from this reprocessing a new deeper
408\,MHz catalogue is being produced. Examination of the reprocessed 
data has revealed
that both MRC\,B1817$-$632
and MRC\,B1754$-$577 appear to be anomalous
north-south sidelobes. However, the other two sources 
(MRC\,B1737$-$602 and MRC\,B2220$-$700) appear to be genuine in the
408\,MHz catalogue. A $b_J=16.75$
magnitude galaxy is located at the position of MRC\,B2220$-$700. 
It is difficult to
know at this stage if MRC\,B1737$-$602 and MRC\,B2220$-$700 
are transient phenomena or form
part of a population of objects whose spectra peak at very low frequency.

\subsection{Resolved Sources}

In the current release of the SUMSS catalogue about 10 per cent of sources
are found to be resolved. This fraction varies from 25 per cent at
\(\delta=-88^\circ\) to 2 per cent at \(\delta=-32^\circ\) where the beam
area is almost doubled. Almost all resolved sources are extended in only 
one direction; only 1 per cent of sources are resolved in both axes. 

To check the accuracy of deconvolved source sizes in the SUMSS catalogue,
sizes and position angles of resolved sources were compared with the sizes
of resolved sources in the NVSS catalogue. Almost all sources found to be
resolved in SUMSS were also resolved in the NVSS; the small fraction that
were not were attributed to the effects described below.

A comparison of the source widths in SUMSS and NVSS is shown in
Figure~\ref{resolvedmajor}. For about half of the sources the
major axis widths agree well
between the two surveys. However, there is a substantial group of sources which
has a larger major axis in SUMSS. 
Many of these sources seem to
be fitted accurately by {\sc vsad} and probably form a class of objects in which much of
their structure has been resolved out in the NVSS images, but is preserved in
the SUMSS survey. This is because of the continuous 
UV coverage of the MOST \citep{bock99}.
Visual inspection of other sources
reveals some of them are poorly fitted by the {\sc vsad} program,
as illustrated in Figure~\ref{fitting}. These sources are
all genuinely resolved close doubles but {\sc vsad} has fitted a gaussian of
major axis much longer than the source. 
 Figure~\ref{resolvedpa}
shows the distribution of position angles between sources in the two
surveys, these are generally in reasonable agreement given the 
very different raw beamshapes in the two surveys.

%possible
%variation in source structure between the different frequencies of the two
%surveys.

\begin{figure}
\includegraphics[width=\linewidth]{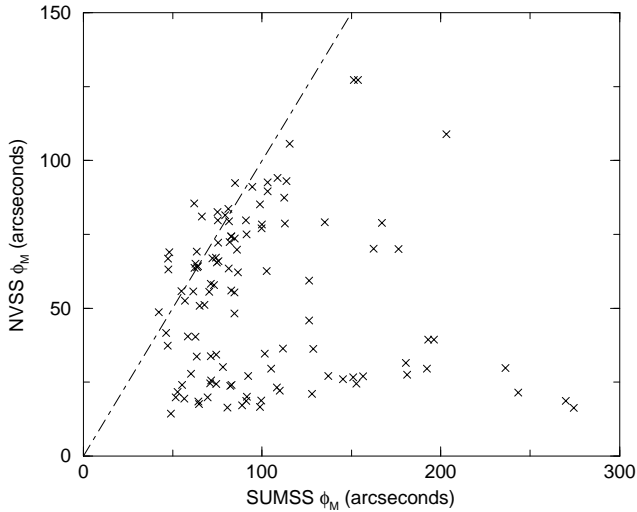}
\caption{The distribution of major axis lengths in SUMSS and NVSS. There is a
  class of SUMSS sources with large angular sizes 
  which appear to be barely resolved
  in NVSS. This is due to a mixture of 
  fitting problems in {\sc vsad} and the increased surface brightness sensitivity
  of the MOST. A small number of sources which are fitted as one elongated
  gaussian in SUMSS are fitted as two separate unresolved
  components in the NVSS, this is because of the larger MOST beam.}
\label{resolvedmajor}
\end{figure}

\begin{figure}
\includegraphics[width=\linewidth]{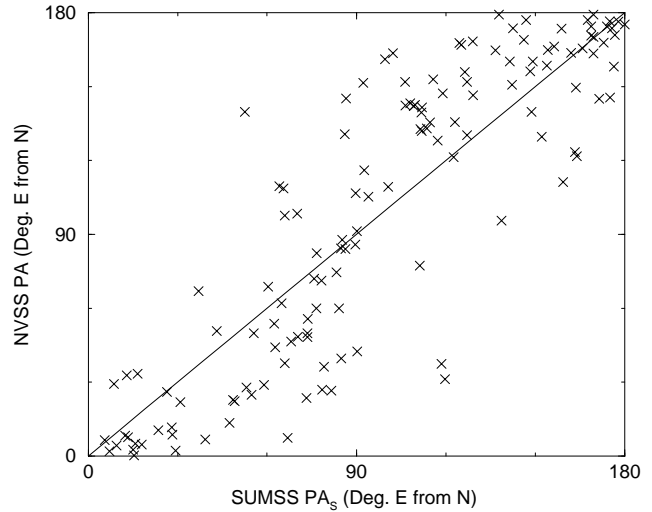}
\caption{The distribution of position angles of resolved sources in SUMSS
  and NVSS. There is a general agreement in position angle between the two
  surveys. Sources with position angles close to \(0^\circ\) in one survey can
  sometimes be found close to \(180^\circ\) in the other, this explains the
  outlying points found on the top left and bottom right corners of the
  plot.}
\label{resolvedpa}
\end{figure} 

\subsection{Source Counts}

Figure~\ref{lognlogs} shows the differential
source counts in the southern part of the
SUMSS catalogue
normalised to a euclidean universe. The source counts shown here are
plotted using peak amplitudes and are used to show the effectiveness of
the decision tree. The true source counts at 843\,MHz should resemble those
shown here fairly closely as only 10 per cent of sources in the catalogue
are resolved. Also, no account is made here for the decrease in source
density close to brighter sources.
A more thorough source count determination will follow in a later paper.

The effect of the decision tree is quite pronounced in this plot. Artefacts
cause the source counts to tail upwards below about 10\,mJy\,beam$^{-1}$. The decision
tree is affecting the counts below this level, causing the counts to
flatten to the same slope as that above 10\,mJy\,beam$^{-1}$. 
This suggests that the
decision tree is doing an excellent job of removing artefacts. Users should
note that the decision tree is having its greatest effect below 
10\,mJy\,beam$^{-1}$, so above
this level the catalogue is mostly unaltered from its raw state.

\begin{figure}
\includegraphics[width=\linewidth,height=7cm]{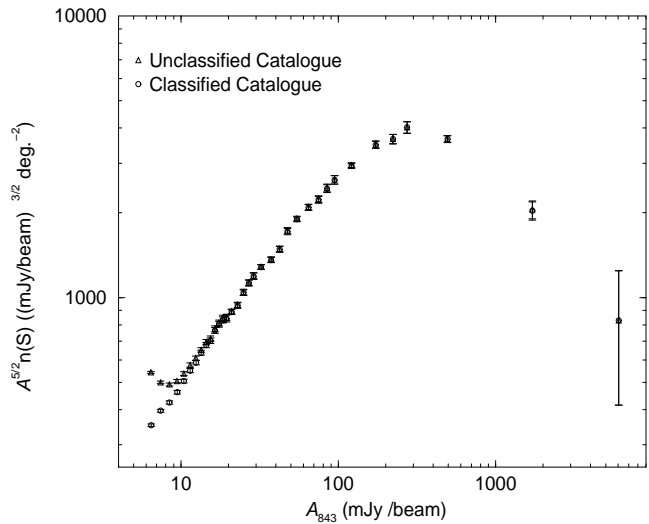}
\caption{The normalised source counts derived from peak amplitude for
  the southern part of the catalogue. As most sources in the survey are
  unresolved this determination should closely resemble the counts for
  total flux density. The counts in the catalogue before
  and after application of the decision tree are presented. Artefacts are
  affecting the decision tree at flux density levels below 
  10\,mJy\,beam$^{-1}$ and the
  decision tree is restoring the counts to the slope which is expected in
  this region.}
\label{lognlogs}
\end{figure}

\section{Summary}

We have created a catalogue of 107,765 sources over the
3500\,deg$^2$ of the SUMSS survey currently available. It is expected
the survey will be complete by the end of 2003. Until then 
we will endeavour to update
the catalogue as new mosaics are released. The survey is currently
progressing at the rate of $1500\,\rm{deg}^2$ per year and we expect the
coverage of the catalogue to increase at a similar rate. As coverage gaps
in the survey are filled positions and flux
densities at the edges of currently released regions may change due to the
increased sensitivity from overlapping fields.
We will maintain backups
of each release of the catalogue so that users may continue to have access
to versions of the catalogue they may previously have used.

We believe the southern catalogue to be 100 per cent complete above $\sim$8\,mJy
and the northern catalogue to be complete above $\sim$18\,mJy. Below these flux
densities, confusion and the decision tree have affected the completeness,
but we expect the catalogue to be highly reliable. 
The catalogue has a source density of 32\,deg$^{-2}$ 
above 6\,mJy\,beam$^{-1}$ and 21\,deg$^{-2}$ above 
10\,mJy\,beam$^{-1}$. Users
should note that some sources (especially those which are resolved) may have
positions and flux densities outside the quoted uncertainties. Users are
encouraged to check the $4.3^\circ \times 4.3^\circ$ mosaics if in doubt.

The catalogue is publicly available and the current version
as well as future updates can be found at 
www.astrop.physics.usyd.edu.au/sumsscat/.

\section*{Acknowledgments}
The Molonglo Observatory site manager, Duncan
Campbell-Wilson, and the staff, Jeff Webb, Michael White and John Barry
are responsible for the smooth operation of the MOST telescope
and the day to day observing program of the SUMSS survey. For this we give
them heartfelt thanks. Bruce McAdam and Tony Turtle provided valuable
information on artefacts in MOST images and deconvolution. Tony Turtle is
also responsible for SUMSS scheduling. The SUMSS survey
is dedicated to Michael Large whose expertise and vision made the project
possible. We would also like to thank the referee J. Condon for some
useful suggestions.
The MOST is operated with the support of the Australian Research 
Council 
and the Science Foundation for
Physics within the University of Sydney.

\end{twocolumn}
\label{lastpage}
\end{document}